\documentclass{aa}  
\usepackage{graphicx}
\usepackage{amssymb}
\usepackage{amsmath}
\usepackage{txfonts}
\usepackage{gensymb}
\usepackage{lscape}
\usepackage{multicol}
\usepackage{subfigure}
\usepackage{soul}
\bibpunct{(}{)}{;}{a}{}{,}
  
\def\phs{\phantom{$-$}}   
 
\def\phn{\phantom{0}}
\def\phnneg{\phantom{$-$}}
\def\nodata{\phs$\cdots$}
\authorrunning{Magee et al.}
\titlerunning{MC LC}

\begin{document} 
            
    \title{Modelling the early time behaviour of type Ia supernovae:} 
    \subtitle{effects of the $^{56}$Ni distribution}
	\author{M. R. Magee \inst{1}
            \and
            S. A. Sim \inst{1}
            \and
            R. Kotak \inst{1,2}
            \and
            W. E. Kerzendorf \inst{3}
            }

	\institute{Astrophysics Research Centre, School of Mathematics and Physics, Queen's University Belfast, Belfast, BT7 1NN, UK \\ 				(\email{mmagee37@qub.ac.uk} \label{inst1})
    \and Tuorla Observatory, Department of Physics and Astronomy, University of Turku, V\"ais\"al\"antie 20, FI-21500 Piikki\"o, Finland \label{inst2}
    \and European Southern Observatory, Karl-Schwarzschild-Str.~2, 85748 Garching bei M\"{u}nchen, Germany \label{inst3}
		}

   \date{Received -	- -; accepted - - - }

% \abstract{}{}{}{}{} 
% 5 {} token are mandatory
 
  \abstract{Recent studies have demonstrated the diversity in type Ia supernovae (SNe Ia) at early times and highlighted a need for a better understanding of the explosion physics as manifested by observations soon after explosion. To this end, 
 we present a Monte Carlo code designed to model the light curves of radioactively driven, hydrogen-free transients from explosion to approximately maximum light. In this initial study, we have used a parametrised description of the ejecta in SNe Ia, and performed a parameter study of the effects of the $^{56}$Ni distribution on the observed colours and light curves for a fixed $^{56}$Ni mass of 0.6\,$M_\odot$. For a given  density profile, we find that models with $^{56}$Ni extending throughout the entirety of the ejecta are typically brighter and bluer shortly after explosion. Additionally, the shape of the density profile itself also plays an important role in determining the shape, rise time, and colours of observed light curves. We find that the multi-band light curves of at least one SNe Ia (SN~2009ig) are inconsistent with less extended $^{56}$Ni distributions, but show good agreement with models that incorporate $^{56}$Ni throughout the entire ejecta. We further demonstrate that comparisons with full $UVOIR$ colour light curves are powerful tools in discriminating various $^{56}$Ni distributions, and hence explosion models. 
}
   \maketitle
%
%________________________________________________________________

\section{Introduction}
\label{sect:intro}

Modern optical transient surveys have led to a wealth of SN discoveries at increasingly early times. As the number of SNe Ia discoveries grows, their diversity becomes ever more apparent - despite being once thought of as a relatively homogeneous group. Particular attention has been paid to epochs shortly after explosion, as they probe the outermost layers of the ejecta and provide information on the progenitor not available at later epochs \cite[e.g.][]{arnett--law, riess--99a, 11fe--nature}. 

\par

A sample of 18 low redshift SNe Ia considered by \cite{firth--sneia--rise} showed significant variation in the rise time to maximum light, ranging from $\sim16$ to 25 days, and some SNe rising more sharply than others. Based on the work of \cite{piro-nakar-2013,piro-nakar-2014}, this variation was interpreted as being due to differences in the distribution of $^{56}$Ni within the SN ejecta. \cite{piro-nakar-2013} show how shallow $^{56}$Ni distributions (i.e. closer to the ejecta surface) lead to shallower rises, and \cite{piro-nakar-2014} apply this study to observations of three SNe Ia shortly after explosion. There is also clear diversity in the observed colours of SNe Ia before and around maximum light \citep{maeda--2011,cartier-2011}. Recent multi-dimensional explosion simulations have argued that SNe Ia may be highly asymmetric \cite[e.g. ][]{livne--2005,kuhlen--2006,kasen--09}. Based on observations of velocity shifts in late-phase nebular spectra, \cite{maeda--2010a} and \cite{maeda--2010b} also argued that SNe Ia may result from asymmetric explosions. The colour differences of \cite{maeda--2011} were found to be correlated with these velocity shifts, therefore \cite{maeda--2011} and \cite{cartier-2011} interpret the observed diversity as resulting from asymmetric explosions. Although some degree of asymmetry has been argued in SNe Ia explosions, the effect of different $^{56}$Ni distributions on the rise time, and in particular colours, has not been fully quantified for even the spherically symmetric case. The purpose of this work is to develop and exploit light curve models that can address this topic and provide physical links between model parameters and observations of SNe Ia.

\par

Analytical work by \cite{arnett--law} modelled the early light curves of SNe with a constant, grey opacity. Subsequent numerical work extended this to variable, grey opacities \citep{cappellaro--97, bersten--2011, morozova-15}.  Further advancements came from incorporating realistic treatments of the plasma state and physics (making as few assumptions as possible), multi-dimensionality, and non-grey opacities \citep{hoeflich-1995a,hoeflich--2003a,stella--98,stella--06,kasen-06b,artis,hillier--2012,vanrossum--2012}. We aim to bridge the gap between these simple and detailed approaches. We present a Monte Carlo radiative transfer code, TURTLS\footnote{TURTLS: The Use of Radiative Transfer for Light curves of Supernovae}, that combines these advantages: it is fast and flexible, like the fixed and grey opacity models, but also implements realistic descriptions of the most important physical processes at the epochs of interest. By including realistic time and frequency dependent opacities, our approach allows us to compute band-limited light curves for any $UVOIR$ filter. This substantially increases the utility of our calculations for direct interpretation of data compared to simpler treatments. The flexibility afforded by our approach coupled with the relatively short ($\sim$dozens of CPU hours) run times facilitates a systematic exploration of the parameter space. It can therefore be used to explore the early light curves and colours of radioactively driven SNe. 

\par

In \S\ref{sect:mcmethod} we describe our code. \S\ref{sect:converge} presents the results of convergence studies and tests of our sensitivity to various approximations made, while \S\ref{sect:comp} presents a comparison to existing codes in the literature for the well-studied W7 \citep{nomoto-w7} explosion model. \S\ref{sect:construct_models} presents a set of models for which light curves are simulated, as discussed in \S\ref{sec:model_lightcurves}. We focus on the effects of the $^{56}$Ni distribution, followed by the density profile. In \S\ref{sect:grey} we demonstrate the importance of a non-grey opacity in determining the light curves. We test the importance of the amount of surface $^{56}$Ni in \S\ref{sec:surface_ni} and quantify the rising phase in \S\ref{sec:rise_index}. \S\ref{sec:09ig} provides a comparison between our models and observations of a SNe Ia. Finally, we present our conclusions in \S\ref{sect:conclusions}.

%
%__________________________________________________________________

\section{Monte Carlo method}
\label{sect:mcmethod}

\subsection{General overview}
\label{subsect:code-overview}
Our Monte Carlo light curve code follows the indivisible energy-packet scheme outlined by \cite{lucy-05} and previous studies \citep{abbott--lucy--85,lucy--abott--93,mazzali--lucy--93,lucy-02,lucy-03}, applied in one dimension. Throughout the following, we adopt the naming convention of \cite{lucy-05}; $\gamma$-packets represent bundles of $\gamma$-ray photons; radiation-packets (r-packets) represent bundles of $UVOIR$ photons; $z, z_1, z_2$, etc. are independent, random numbers. 

\par

Packets represent discrete monochromatic bundles of photons. They are injected into the model region and their propagation followed, during which they may undergo electron scattering or absorption and re-emission by the ions and atoms in the ejecta. The number of photons and frequencies represented by a packet may change during the simulation; however, energy is always conserved in the fluid frame, and packets are neither created nor destroyed during interactions with the ejecta. Radiative equilibrium and conservation of the total energy are therefore enforced throughout the simulation. Observed light curves are created by binning emerging packets in frequency and time. 

\par

Throughout our simulations, packet properties such as propagation direction, energy, etc. are followed in the observer frame, and transformed to and from the fluid frame when appropriate, through a first-order Doppler correction. 

\subsection{Model set-up}
\label{subsect:model-setup}

In our implementation, the form of the SN ejecta is entirely free, and is assumed to be spherically symmetric. The ejecta is defined by a series of zones, each having the following properties upon input: inner and outer velocity boundaries; density at some reference time, $t_0$; $^{56}$Ni mass fraction and composition at $t_0$. 

\par

The temperature of each zone at the start of the simulation, $t_{\rm{s}}$, is determined by the local heating that has occurred due to the decay of $^{56}$Ni since the explosion. Following \cite{lucy-05}, we determine the radiation energy density in each zone as:
\begin{equation}
U_{\rm{R}} = 
\left(
\frac{t_{\rm{Ni}}}{t_{\rm{s}}} - 
\left[
\left( 
1 + \frac{t_{\rm{Ni}}}{t_{\rm{s}}}
\right)
\exp\left(\frac{-t_{\rm{s}}}{t_{\rm{Ni}}}\right)
\right]
\right)
\frac{\chi~\rho~E_{\rm{{Ni}}}}{m_{\rm{Ni}}},
\end{equation}
where $t_{\rm{Ni}}$ is the decay time of $^{56}$Ni (8.8 days), $\chi$ is the initial mass fraction of $^{56}$Ni in that zone, $\rho$ is the density of that zone at $t_{\rm{s}}$, $E_{\rm{Ni}}$ is the energy emitted by the decay of a $^{56}$Ni atom (1.728 MeV), and $m_{\rm{Ni}}$ is the mass of a $^{56}$Ni atom (9.3 $\times 10^{-26}$~kg). The mean intensity of the radiation field in each zone is given by:
\begin{equation}
J = \frac{U_{\rm{R}}c}{4\pi},
\end{equation}
and the radiation temperature of each zone by:
\begin{equation}
\label{temperature}
T_{\rm{R}}^{4} = \frac{\pi J}{\sigma_{SB}}.
\end{equation}

For zones without $^{56}$Ni at the start of the simulation, we set an initial temperature of 1000~K. We treated 1000~K as a minimum temperature in each zone throughout the entire simulation. We tested other values for a minimum temperature (up to 5000~K) but found that this did not have a noticeable effect on the resultant light curves. For the epochs considered in this work, we do not expect the outer temperature to be significantly lower.

\subsection{Packet initialisation}
\label{subsect:packet-init}

With the model region constructed, packets must be initialised with the following properties: time of injection, position, direction of propagation, and energy. Following the $^{56}$Ni decay chain, $^{56}$Ni~$\rightarrow$~$^{56}$Co~$\rightarrow$~$^{56}$Fe, $\gamma$-packets are injected and randomly assigned an emitting species (proportional to their probabilities) of either a $^{56}$Ni or $^{56}$Co nucleus. The $^{56}$Ni decay chain is assumed to be the dominant form of energy production. Other decay chains are therefore not implemented currently, but could be included in future work. The injection time for $^{56}$Ni decays is given by $t_{\gamma}$ = --$\ln{z} \times t_{\rm{Ni}}$. For $^{56}$Co decays, the time of injection is given by $t_{\gamma}$ = --$\ln{z_1} \times t_{\rm{Ni}}$ --$\ln{z_2} \times t_{\rm{Co}}$, where $t_{\rm{Co}}$ is the decay time of $^{56}$Co. Following from \cite{lucy-05}, $\gamma$-packets injected before the start time of the simulation are converted to r-packets, and given an injection time equal to this start time. Work done by packets on the ejecta during this time is also accounted for. 

\par

The radial position of the packet at the injection time is determined by sampling the $^{56}$Ni distribution within the SN ejecta. The distribution of propagation directions within the fluid frame is assumed to be isotropic. Each packet is assigned a random propagation direction, $\mu_{\rm{f}} = 2z - 1$, which is then transformed into the observer frame following \cite{castor--72}. The energy of each packet is given by discretising the total energy emitted by the SN in the fluid frame among all packets.

\par

Packets that have converted from $\gamma$- to r-packets by the start of the simulation also require an initial frequency. The fluid frame frequency is selected by sampling the Planck function at the temperature appropriate for that packet's zone, following from \cite{carter-cashwell-75} and \cite{bjorkman-wood-01}. 

\subsection{Packet propagation}
\label{subsect:packet-prop}

With all packets initialised, the simulation can begin. The simulation operates by propagating packets in logarithmically spaced time intervals between chosen start, $t_{\rm{s}}$, and end, $t_{\rm{e}}$, times. 

\par

Once a packet has been injected we simulate its random walk by calculating four time intervals: time for the packet to redshift into the next frequency bin ($t_{\rm{f}}$), time for the packet to reach a zone boundary ($t_{\rm{b}}$), time for the packet to reach an interaction point ($t_{\rm{i}}$), and time until the end of the current time step (t$_{\rm{nts}}$). We propagate each packet until it reaches the first of these four events to occur and perform the event. New time intervals are then calculated and the procedure repeated, if necessary. This is performed for all packets in all time steps until the chosen end of the simulation is reached. 

\par

In the following, we describe each interval in more detail. We begin by describing numerical events in \S\ref{subsect:numerical}, followed by physical events in \S\ref{subsect:packet-interactions}.

\subsubsection{Numerical events}
\label{subsect:numerical}
As packets propagate, their fluid frame frequency is constantly redshifted. If the time to the next frequency bin is shortest, the packet is propagated to this point, the fluid frame frequency updated, and new time intervals calculated.

\par

The trajectory of the packet is followed from its current position until it intersects either the inner or outer boundary of the zone - depending on the direction of travel. If the time to a boundary is shortest, the packet is propagated to the appropriate boundary and new time intervals are calculated. If a packet is propagated to the outer boundary of the final zone, it has escaped the simulation. See \S\ref{subsect:construct-lightcurves} for further details. 

\par

If the beginning of the next time step is the next event to occur, the packet is propagated along its current trajectory until the end of the current time step. Once the next time step has begun, the process of calculating new time intervals begins again.

\subsubsection{Physical events}
\label{subsect:packet-interactions}

The final interval calculated is the time until a packet reaches a randomly selected optical depth, given by:

\begin{equation}
t_{\rm{i}} = \frac{ -\ln{z}}{c\rho\kappa},
\end{equation}
where $\rho$ is the density of the packet's zone, and $\kappa$ is the opacity. 

\par

\begin{figure}
\centering
\includegraphics[width=\columnwidth]{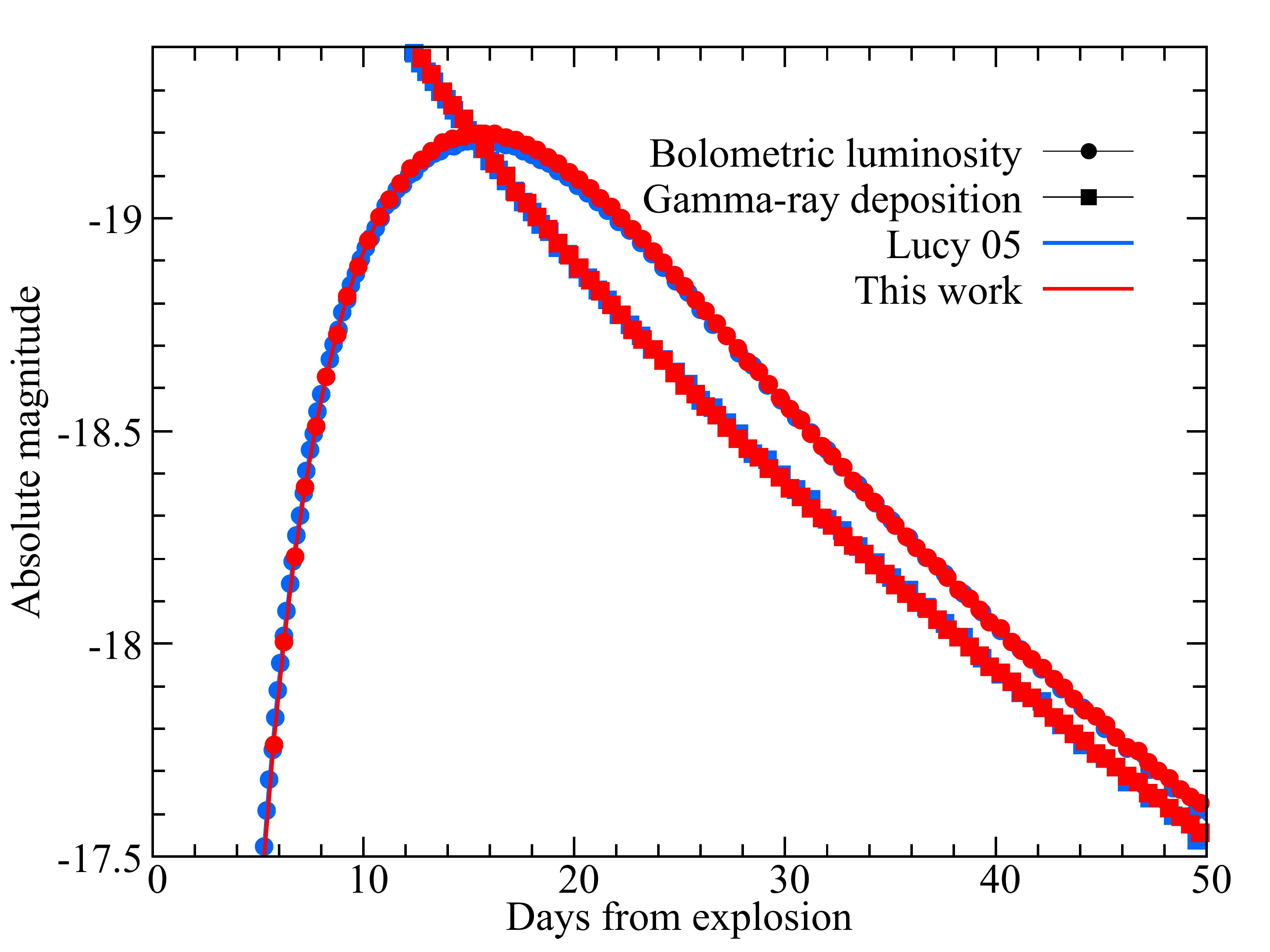}
\caption{Bolometric light curve and $\gamma$-ray deposition curve for the model presented by \cite{lucy-05} compared to a calculation performed in this work. Our code reproduces the results of \cite{lucy-05}. }
\label{fig:lucy05_comp}
\centering
\end{figure}

For $\gamma$-packets, we used a fixed grey opacity of $\kappa/\rho$ = 0.03~cm$^2$~g$^{-1}$ \citep{ambwani-88}. Despite this approximation, our code is able to reproduce the $\gamma$-ray deposition obtained using the more sophisticated treatment of \cite{lucy-05} (see \S~\ref{sect:comp_lucy},  Fig.~\ref{fig:lucy05_comp}). For a $\gamma$-packet experiencing its first interaction, it is immediately destroyed and re-emitted as an r-packet. 

\par

For r-packets, we used TARDIS \citep{tardis} to calculate Thomson scattering opacities and expansion opacities \citep{eastman-pinto-93}. We assumed LTE when determining the ionisation and excitation levels. At the end of each time step, the zone densities and temperatures, and time since explosion are input into TARDIS to calculate the opacities. We also took into account the change in composition that results from the decay of $^{56}$Ni $\rightarrow$ $^{56}$Co $\rightarrow$ $^{56}$Fe. 

\par

The use of expansion opacities represents a significant simplification in the treatment of photon-matter interactions. As our atomic dataset includes $4.5 \times 10^5$ lines,  treating each line individually would be computationally prohibitive. In the expansion opacity formalism, lines are combined into opacity bins in discrete frequency intervals. The expansion opacity in the frequency interval $\nu + \Delta\nu$ is therefore the contribution from all lines within the given interval, and calculated as:

\begin{equation}
\label{eqn:exp_opacity}
\kappa_{\mathrm{exp}}\left(\nu\right) = \frac{\nu}{\Delta \nu} \frac{1}{c t_{\rm{exp}}} \sum_j \left(1 - \exp\left(-\tau_{\mathrm{S},j}\right)\right),
\end{equation}
where $t_{\rm{exp}}$ is the time since explosion, and $\tau_{\rm{S,j}}$ is the Sobolev optical depth for the line $j$, given by:

\begin{equation}
\tau_S = \dfrac{\pi e^{2}}{m_{e} c}f \lambda_{\rm{lu}} t_{\rm{exp}}\
        n_{\rm{l}} \left(1-\dfrac{g_{\rm{l}}n_{\rm{u}}}{g_{\rm{u}}n_{\rm{l}}}\right),
\end{equation}
where $f$ is the absorption oscillator strength of the transition, as in \cite{tardis} 

\par

If an interaction is the next event, the packet is propagated to the interaction point and a new fluid frame direction randomly selected. Energy is always conserved in the fluid frame during interactions. The form of interaction (electron-scattering or absorption) is chosen randomly at the point of interaction, in proportion to their probabilities. If a packet is electron-scattered, we also conserve frequency in the fluid frame.

\par

For packets that experience absorption, we use the two-level atom approach (TLA) outlined by \cite{kasen-06b}, which greatly reduces the computational demands of the simulation. In real systems, when an atom absorbs a photon through a line transition, there may be multiple transitions through which that photon could be re-emitted. In the TLA approach, once packets are absorbed, they are immediately re-emitted with either the same frequency or a new frequency chosen by randomly sampling the local thermal emissivity, $S(\nu_{\rm{i}})$, given by:

\begin{equation}
\label{eqn:source}
S\left(\nu_{\rm{i}}\right) = B\left(\nu_{\rm{i}}\right)\kappa\left(\nu_{\rm{i}}\right),
\end{equation}
where $B\left(\nu_{\rm{i}}\right)$ is the Planck function, and $\kappa\left(\nu_{\rm{i}}\right)$ is the expansion opacity. The probability of redistribution, as opposed to coherent line scattering, is given by the redistribution parameter, $\epsilon$. In principle, $\epsilon$ is unique for each line, however, \cite{kasen-06b} show that a single value close to unity for all lines is sufficient to reproduce a more detailed treatment of fluorescence. Throughout our simulations, we fix $\epsilon = 0.9$. Varying $\epsilon$ is discussed in \S\ref{fluorescence}. Once a packet has finished the interaction process, new time intervals are calculated. 

\subsection{Updates to the plasma state}
\label{subsect:plasma}
Once all packets have propagated through the current time step, the density, temperature, and source function for each zone are updated. Following from \cite{mazzali--lucy--93} and \cite{long--knigge--02}, we use a Monte Carlo estimator to determine the mean intensity in each zone, given by:

\begin{equation}
J_{\rm{est}} = \frac{1}{4\pi\Delta tV}\sum{ElD_{\rm{\mu}}},
\end{equation}
where $\Delta  t$ is the time step duration, $V$ is the volume of the zone, $E$ is the energy of the packet, and $l$ is the distance travelled by the packet during the current time step. The summation is performed over all packets that have travelled inside the zone during the current time step. We then calculate new temperatures using Eqn.~\ref{temperature}, new opacities using TARDIS, and update the source function (Eqn.~\ref{eqn:source}) in each zone. 

\subsection{Constructing light curves}
\label{subsect:construct-lightcurves}
 
As packets escape the outer grid zone with an escape time $t_{\rm{n}}$, they are perceived by a distant observer to have been emitted at an observed time $\tau_{\rm{n}}~=~t_{\rm{n}} - \mu R_{\rm{max}}$/c \citep{lucy-05}. R-packets that have escaped the model domain are binned in observed time and frequency. Frequency bins are then convolved with the desired set of filter functions to construct the model light curves.

%
%______________________________________________________________

\section{Convergence and sensitivity tests}
\label{sect:converge}

In this section, we describe the tests conducted to assess the robustness of our results. All calculations described in this section use the density and composition from the W7 explosion model \citep{nomoto-w7}.

\subsection{Number of packets}
\label{packet-no}

As discussed in \S\ref{subsect:plasma}, we use a Monte Carlo estimator to determine the temperature of each zone. Therefore, the number of packets will affect not only the noise present in the light curve but also the opacities used in the simulation. We find that, even with a relatively small number of packets ($10^5$), noise in the temperature profile does not have a noticeable effect on the light curves beginning $\sim$3 days post-explosion.

\par

The effect of a small number of packets is most pronounced at early times ($\lesssim2.5$ days post explosion) for models that do not have $^{56}$Ni extending to the surface of the ejecta. In these models, the diffusion of packets into the outer regions of the ejecta can cause relatively large fluctuations in the temperature across zones, as individual zones may contain only a handful of packets. Nevertheless, sufficiently good statistics can be achieved by selecting an appropriate time span for the simulation.

\subsection{Number of time steps}
\label{timestep-no}

After each time step, the plasma state is updated with new temperatures and source functions in each zone, and new opacities are calculated. The number of time steps may affect the light curve if they are too few to accurately represent the smooth evolution of the SN ejecta. We test models with 50, 100, 250, and 500 time steps logarithmically spaced between 1.5 and 30 days post explosion. We found that even with a relatively small number of steps, light curves showed little variation. Typically the largest difference between these light curves is $\lesssim$0.05 mag for models with more than 100 time steps. We use 250 time steps, as a compromise between accurately capturing the light curve evolution and the computational requirement. 

\subsection{Number of frequency bins}
\label{freq-no}

The use of expansion opacities requires that lines are collected into discrete frequency bins. We calculate models for 1000, 5000, and 10,000 bins ranging from $10^{14} - 10^{16}$ Hz, and find that increasing the number of frequency bins beyond 1000 has a negligible effect on the resultant light curves ($\lesssim0.05$ mag).

\subsection{Atomic dataset}
\label{atomic-data}

Our atomic dataset comprises lines drawn from \cite{kurucz-95}, with a cut in $\log(gf)$ applied to limit the number of weak lines included, and hence the computational time requirement, as in \cite{artis}. We find that a cut of $\log(gf)\geq-2$ typically produces brighter light curves, particularly in bluer bands. It is unsurprising that a smaller atomic data set would result in brighter early light curves, given that a reduction in the number of lines will result in a reduced opacity. That the effect is most prominent at short wavelengths highlights the importance of the weaker lines due to iron group elements. The difference between $-3$ and $-5$ cuts is far less pronounced but follows the same trend. Including even weaker lines ($\log(gf)\geq-20$) does not alter the model light curves \citep{artis}. We therefore use the $\log(gf)\geq-$5 atomic data set.

\par

\cite{artis} also performed tests with atomic data sets that include many more weak lines due to singly- and doubly-ionised \ion{Fe}, \ion{Co}, and \ion{Ni}. These transitions primarily occur at red and NIR wavelengths. \cite{artis} showed that including these lines can produce a noticeable deviation in the NIR bands (particularly $J, H$, and $K$) but has little effect on the optical light curves.

\par

As shown by \cite{dessart--2014b}, forbidden line transitions play an important role in cooling from $\sim30$ days post explosion. We note that the atomic dataset used in our simulations includes only permitted line transitions, and therefore do not extend our simulations beyond approximately maximum light.

\subsection{Fluorescence parameter}
\label{fluorescence}

\begin{figure*}[!h]
\centering
\includegraphics[width=\textwidth]{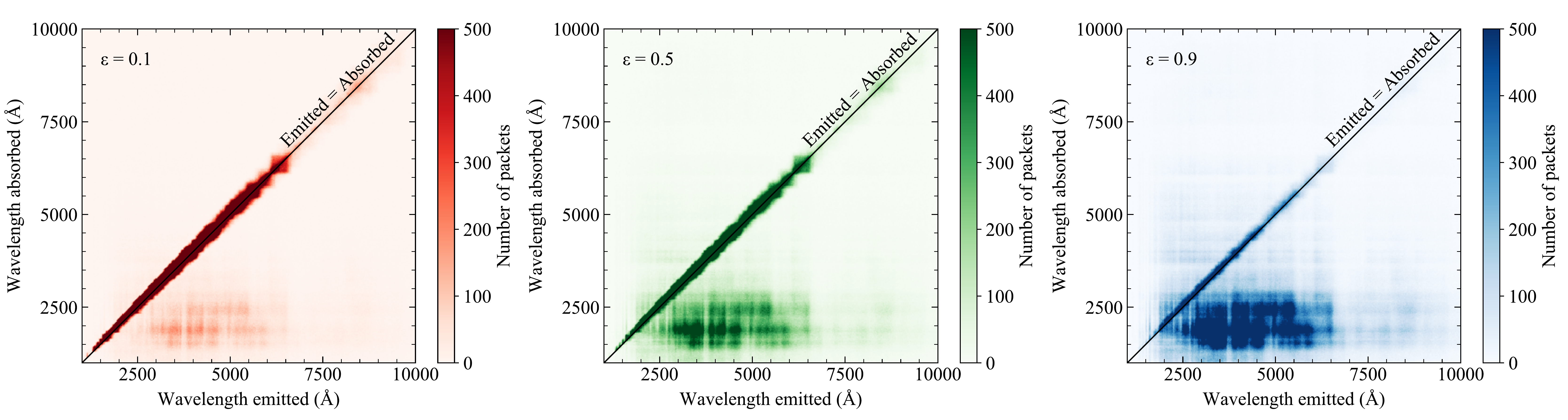}
\caption{Wavelength at which packets are absorbed and re-emitted following their last interaction, and before exiting the simulation. Three values of the redistribution parameter are shown from left to right: scattering dominated ($\epsilon = 0.1$), equal probability for scattering and redistribution ($\epsilon = 0.5$), and redistribution dominated ($\epsilon = 0.9$). 
}
\label{fig:redistribution}
\centering
\end{figure*}

We use the two-level atom approach - defined through the use of the redistribution parameter, $\epsilon$ - to approximate fluorescence \citep{kasen-06b}. The probability that packets will be re-emitted with frequencies sampled from the source function is given by $\epsilon$.

\par

As $\epsilon$ values close to unity have been shown to broadly reproduce the effects of a full fluorescence treatment \citep{kasen-06b}, we adopted $\epsilon = 0.9$ throughout our simulations. Figure~\ref{fig:redistribution} shows the redistribution of packet wavelengths for three models ($\epsilon = 0.1$, 0.5, and 0.9), following the last interaction experienced by each packet, and demonstrates how models with a low $\epsilon$ value (0.1) do not redistribute as effectively following interactions.

%
%______________________________________________________________

\section{Code verification}
\label{sect:comp}
In this section, we test whether our code is consistent with other radiative transfer codes. We first compared our code using models calculated with a grey opacity (\S\ref{sect:comp_lucy}), followed by models calculated with a non-grey opacity (\S\ref{sect:comp_nongrey}).

\par

\subsection{Grey opacity}
\label{sect:comp_lucy}
We compare the simple, grey opacity ($\kappa/\rho$ = 0.1~cm$^{2}$~g$^{-1}$) model presented by \cite{lucy-05} to a calculation performed using our code that also incorporates a fixed, grey opacity. This model has a uniform density structure, with an ejecta mass of 1.39~$M_{\odot}$, a $^{56}$Ni mass of 0.625~$M_{\odot}$, and a maximum velocity of 10$^4$~km~s$^{-1}$. The mass fraction of $^{56}$Ni at the centre of the ejecta (M(r) \textless ~0.5~$M_{\odot}$) is equal to one, and drops linearly to zero at M(r) = 0.75~$M_{\odot}$.

\par

We used a grey opacity for $\gamma$-ray transport \citep[0.03~cm$^{2}$~g$^{-1}$;][]{ambwani-88}, while \cite{lucy-05} perform a more complete treatment, including non-grey opacities and Compton scattering. Figure~\ref{fig:lucy05_comp} shows, however, that our code is able to reproduce the results of \cite{lucy-05}, including the $\gamma$-ray deposition. This suggests that our simplified approach to $\gamma$-ray transport does not have a significant effect on our model light curves.

\par

\subsection{Non-grey opacity}
\label{sect:comp_nongrey}
\begin{figure*}
\centering
\includegraphics[scale=0.32]{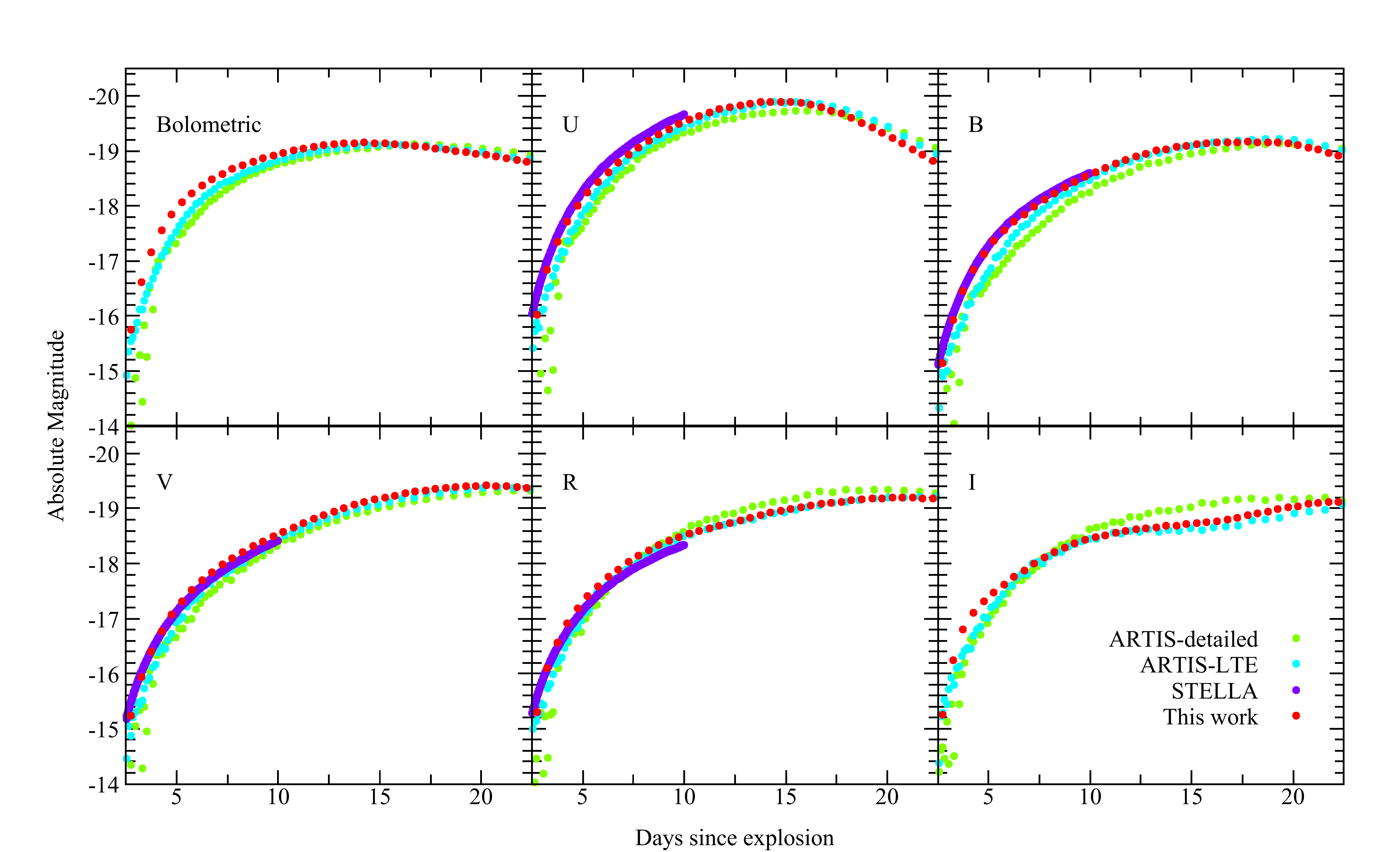}
\caption{Comparison of bolometric and $UBVRI$ light curves for the W7 explosion model generated by different light curve codes. ARTIS and STELLA light curves are from by \cite{artis} and \cite{noebauer-17}, respectively.
}
\label{fig:code_comp}
\centering
\end{figure*}

We compare to other non-grey radiative transfer codes using the well studied W7 explosion model \citep{nomoto-w7}. We use $1\times10^7$ packets, 1000 frequency bins, 250 time steps, and begin the simulation 1.5 days after explosion. 

\par

ARTIS \citep{artis} is a 3D Monte Carlo radiative transfer code for calculating time-dependent supernova spectra. An important difference between our code and ARTIS is that ARTIS does not use the expansion opacity approximation, but treats each line individually. This is a significant improvement over our code; however, ARTIS also requires orders of magnitude greater computational time ($\sim$100 vs. $\sim$5,000 CPU hours for the models presented here). Figure~\ref{fig:code_comp} presents ARTIS light curves calculated using an LTE approximation with a standard atomic data set ($4.1\times10^5$ lines), and a more detailed ionisation treatment with a larger atomic data set ($8.2\times10^6$ lines). This larger atomic data set includes more transitions for singly- and doubly-ionised \ion{Fe}, \ion{Co}, and \ion{Ni}. See \cite{artis} for further details.

\par

Figure~\ref{fig:code_comp} demonstrates that the expansion opacity approximation is able to reproduce the full line treatment implemented in ARTIS. In particular, our model W7 light curves more closely match those of the LTE ARTIS model - this is unsurprising given the LTE assumptions used in calculating our opacities.

\par

Figure~\ref{fig:code_comp} also shows early-phase light curves calculated by \cite{noebauer-17} using STELLA \citep{stella--98,stella--06}, a one-dimensional radiation hydrodynamics code. Similar to our code, LTE is assumed in determining the ionisation and excitation levels in STELLA. STELLA also makes use of a slightly smaller atomic data set, $\sim1.6\times 10^5$ lines, and the expansion opacity approximation. Light curves produced by STELLA also show good agreement with those produced by our code, and ARTIS.

%
%______________________________________________________________

\section{Construction of model density and composition profiles}
\label{sect:construct_models}
Having described the operation of our code and demonstrated that it produces results similar to other codes in the literature (see \S\ref{sect:comp}), we apply it to test the effects of varying the $^{56}$Ni distribution. We describe the set-up of the models presented in this work, while \S\ref{sec:model_lightcurves} presents the light curves.

\begin{table*}
\centering
\caption{Ejecta model parameters and properties}\tabularnewline
\label{tab:model-params}\tabularnewline
\resizebox{\textwidth}{!}{
\begin{tabular}{lllllllllllcc}\hline
\hline
\tabularnewline[-0.25cm]
Model &		Transition 				   & Inner   	& Outer & Scale 	& Bolometric & 		  & $B$	   &	   & $V$ 	    &	    & $B-V$ 	   & $B-V$  \tabularnewline
&	velocity 				   & slope   	& slope & parameter & Rise 	  	 & Peak   & Rise   & Peak  & Rise   & Peak  & t = 2.5\,d & t = $B$ max\tabularnewline\textbf{}
&	v$_{\rm{t}}$ (km~s$^{-1}$) & $\delta$ 	& n 	& s 	& (days)  	 & (mag)  & (days) & (mag) & (days) & (mag) & (mag)    & (mag)\tabularnewline
 
\hline
\hline
\tabularnewline[-0.2cm]
v7500\_d0\_n8\_s3		&	\phn7\,500 	& 0 & \phn8 & \phn\phn3		&	18.7 &	$-$18.88	& 16.9	& $-$19.16 &	23.7 &	$-$19.37 	&	$-$0.04	&	\phnneg0.08	\tabularnewline 
v7500\_d0\_n8\_s9.7		&	\phn7\,500 	& 0 & \phn8 & ~9.7	&	20.0 &	$-$18.89	& 22.6	& $-$19.02 &	24.4 &	$-$19.15 	&	\phnneg0.45&	\phnneg0.10	\tabularnewline 
v7500\_d0\_n8\_s100		&	\phn7\,500 	& 0 & \phn8 & 100	&	19.0 &	$-$18.78	& 13.8	& $-$18.89 &	\nodata	 &	\nodata			&	\phnneg0.43&	 $-$0.24	\tabularnewline 
v7500\_d0\_n12\_s3		&	\phn7\,500 	& 0 & 12 	& \phn\phn3		&	19.1 &	$-$18.83	& 19.9	& $-$19.07 &	\nodata	 &	\nodata 			&	$-$0.19	&	\phnneg0.05	\tabularnewline 
v7500\_d0\_n12\_s9.7	&	\phn7\,500 	& 0 & 12 	& ~9.7	&	20.0 &	$-$18.82	& 24.1	& $-$18.97 & \nodata	 &	\nodata			&	\phnneg0.21&	\phnneg0.05	\tabularnewline 
v7500\_d0\_n12\_s100	&	\phn7\,500 	& 0 & 12 	& 100	&	20.2 &	$-$18.69	& 19.6	& $-$18.77 &	\nodata	 &	\nodata			&	\phnneg0.65&	 $-$0.07	\tabularnewline 
v7500\_d1\_n8\_s3		&	\phn7\,500 	& 1 & \phn8 & \phn\phn3		&	20.1 &	$-$18.83	& 17.9	& $-$19.11 &	23.8 &	$-$19.33 	&	$-$0.08	&	\phnneg0.10	\tabularnewline 
v7500\_d1\_n8\_s9.7		&	\phn7\,500 	& 1 & \phn8 & ~9.7	&	21.4 &	$-$18.85	& \nodata	& \nodata	   &	\nodata    &	\nodata 			&	\phnneg0.41&	\nodata		\tabularnewline 
v7500\_d1\_n8\_s100		&	\phn7\,500 	& 1 & \phn8 & 100	&	20.2 &	$-$18.74	& 14.4	& $-$18.86 &	\nodata    &	\nodata			&	\phnneg0.81&	 $-$0.25	\tabularnewline 
v7500\_d1\_n12\_s3		&	\phn7\,500 	& 1 & 12 	& \phn\phn3		&	19.9 &	$-$18.78	& 20.5	& $-$19.05 &	\nodata    &	\nodata 			&	$-$0.19	&	\phnneg0.06	\tabularnewline 
v7500\_d1\_n12\_s9.7	&	\phn7\,500 	& 1 & 12 	& ~9.7	&	23.2 &	$-$18.79	& 24.4	& $-$18.95 &	\nodata    &	\nodata			&	\phnneg0.30&	\phnneg0.02	\tabularnewline 
v7500\_d1\_n12\_s100	&	\phn7\,500 	& 1 & 12 	& 100	&	20.7 &	$-$18.65	& 20.2	& $-$18.74 &	\nodata    &	\nodata 			&	\phnneg0.76&	 $-$0.09	\tabularnewline 
v12500\_d0\_n8\_s3		&	12\,500 	& 0 & \phn8 & \phn\phn3		&	14.9 &	$-$19.05	& 12.9	& $-$19.27 &	17.8 &	$-$19.66 	&	\phn0.11&	\phnneg0.21	\tabularnewline 
v12500\_d0\_n8\_s9.7	&	12\,500 	& 0 & \phn8 & ~9.7	&	16.3 &	$-$19.11	& 17.2	& $-$19.33 &	19.0 &	$-$19.52 	&	\phnneg0.33&	\phnneg0.15	\tabularnewline 
v12500\_d0\_n8\_s100	&	12\,500 	& 0 & \phn8 & 100	&	16.7 &	$-$19.06	& 18.0	& $-$19.10 &	18.0 &	$-$19.31 	&	\phnneg0.35&	\phnneg0.21	\tabularnewline 
v12500\_d0\_n12\_s3		&	12\,500 	& 0 & 12 	& \phn\phn3	&	15.3 &	$-$19.03	& 14.3	& $-$19.26 &	19.0 &	$-$19.55 	&	$-$0.17	&	\phnneg0.13		\tabularnewline 
v12500\_d0\_n12\_s9.7	&	12\,500 	& 0 & 12 	& ~9.7	&	16.6 &	$-$19.06	& 17.8	& $-$19.28 &	20.4 &	$-$19.43 	&	\phnneg0.14&	\phnneg0.10	\tabularnewline 
v12500\_d0\_n12\_s100	&	12\,500 	& 0 & 12 	& 100	&	17.5 &	$-$18.99	& 18.4	& $-$19.04 &	19.4 &	$-$19.24 	&	\phnneg0.23&	\phnneg0.18	\tabularnewline 
v12500\_d1\_n8\_s3		&	12\,500 	& 1 & \phn8 & \phn\phn3		&	16.4 &	$-$19.00	& 14.2	& $-$19.18 &	18.9 &	$-$19.64 	&	\phn0.08&	\phnneg0.31	\tabularnewline 
v12500\_d1\_n8\_s9.7	&	12\,500 	& 1 & \phn8 & ~9.7	&	17.3 &	$-$19.07	& 18.0	& $-$19.31 &	20.8 &	$-$19.51 	&	\phnneg0.32&	\phnneg0.14	\tabularnewline 
v12500\_d1\_n8\_s100	&	12\,500 	& 1 & \phn8 & 100	&	17.6 &	$-$19.01	& 19.2	& $-$19.05 &	20.0 &	$-$19.28 	&	\phnneg0.47&	\phnneg0.23	\tabularnewline 
v12500\_d1\_n12\_s3		&	12\,500 	& 1 & 12 	& \phn\phn3	&	16.6 &	$-$18.98	& 15.0	& $-$19.19 &	20.3 &	$-$19.55 	&	$-$0.17 &	\phnneg0.18		\tabularnewline 
v12500\_d1\_n12\_s9.7	&	12\,500 	& 1 & 12 	& ~9.7	&	17.9 &	$-$19.02	& 18.7	& $-$19.26 &	22.0 &	$-$19.45 	&	\phnneg0.17&	\phnneg0.10	\tabularnewline 
v12500\_d1\_n12\_s100	&	12\,500 	& 1 & 12 	& 100	&	18.1 &	$-$18.95	& 16.3	& $-$18.99 &	21.2 &	$-$19.22 	&	\phnneg0.39&	\phnneg0.02	\tabularnewline 
\hline
\end{tabular}
}
\tablefoot{Parameters of the artificial density profiles. We fit each light curve with a sixth order polynomial to determine the rise time to peak and peak absolute magnitude for the bolometric, $B-$, and $V-$band light curves. We also use the fits to determine the $B-V$ colour for t = 5 days and at $B-$band maximum. We note that our polynomial fit to determine the colour at 2.5 days is performed for the light curve between two and ten days. For all other cases the fit is performed between 10 and 25 days. Times of maximum that occur later than $\sim$24.5 days are neglected. 
}
\end{table*}

Our code requires the density and composition of the SN ejecta as input, both of which are freely defined by the user. Following \cite{botyanszki-2017}, we parametrise the density profile of the ejecta as a broken power law. This produces an ejecta with a shallow inner region, and a more steeply declining outer region. The density at velocity $v$ is given as:
\begin{equation}
\rho(v) =\left\{
	\begin{array}{lr}
     \rm \rho_0~~\left(v/v_{\rm{t}}\right)^{-\delta} & \rm ~v \leq v_{\rm{t}}~ \\  
     \rm \rho_0~~\left(v/v_{\rm{t}}\right)^{-n} & \rm ~v > v_{\rm{t}}, \\
	\end{array} 
\right.
\end{equation}
where v$_{\rm{t}}$ gives the velocity boundary between the two regions, $\delta$ gives the slope of the inner region, $n$ gives the slope of the outer region, and the reference density, $\rho_0$, is given by:
\begin{equation}
\rm \rho_0 = \frac{M_{ej}}{4 \pi \left(v_{\rm{t}} t_{exp}\right)^3} \left[\frac{1}{3 - \delta} + \frac{1}{n - 3}\right]^{-1},
\end{equation}
where $\delta$ \textless 3, n \textgreater 3, M$_{\rm{ej}}$ is the ejecta mass, and t$_{\rm{exp}}$ is the time since explosion \citep{botyanszki-2017}. In order to test only the effects of the $^{56}$Ni distribution, we fix the mass and maximum velocity of the ejecta to be 1.4~$M_{\odot}$ and 30\,000~km~s$^{-1}$, respectively, for all density profiles discussed in this section. 
\par
Parameters of the models used in this study are given in Table~\ref{tab:model-params}. These density profiles were constructed such that they broadly span the range predicted by various explosion models, such as pure deflagrations \citep{nomoto-w7,fink-2014}, deflagration-to-detonation transitions \citep{seitenzahl-2013}, and the violent merger of two WDs \citep{pakmor-2012}.

\par

\begin{figure*}[!t]
\centering
\includegraphics[scale=0.27]{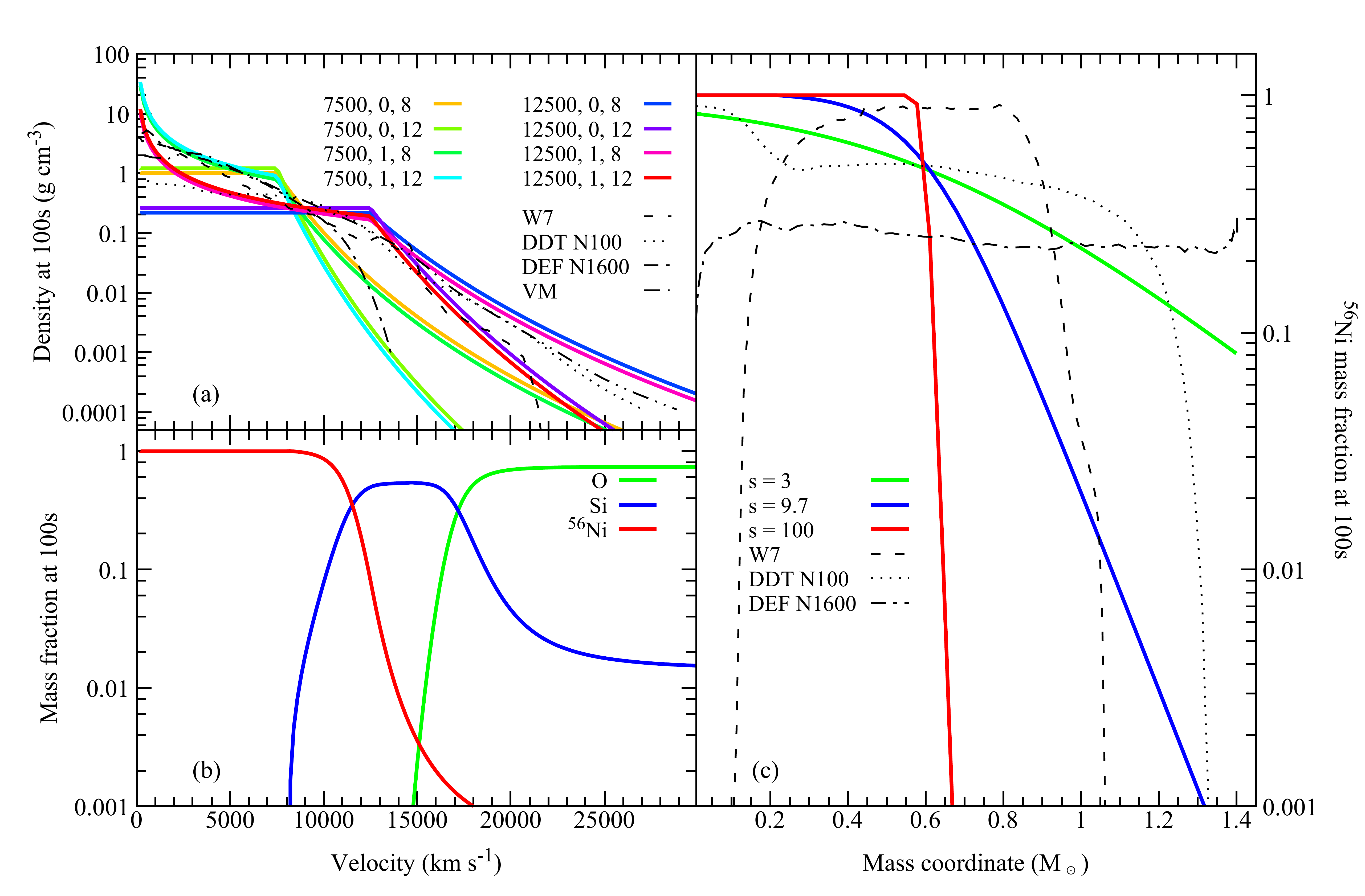}
\caption{\textbf{Ejecta models used in this work.} (a) Density profiles for our model sequence. Transition velocity, delta, and n parameters are given for each model. Densities for a sample of explosion models are shown for comparison: W7 \citep{nomoto-w7}, DDT N100 \citep{seitenzahl-2013}, DEF N1600 \citep{fink-2014}, VM \citep{pakmor-2012}. (b) Illustrative composition profile for one model (v12500\_d0\_n8\_s9.7). The three zone structure is clearly demonstrated, with representative species shown. (c) $^{56}$Ni distributions for one density profile (v12500\_d0\_n8). $^{56}$Ni distributions for a sample of explosion models are also shown. }
\label{fig:model-profiles}
\centering
\end{figure*}

We have constructed $^{56}$Ni distributions such that $^{56}$Ni is concentrated towards the inner ejecta, and extends outwards to varying degrees - approximately following predictions by explosion models for SNe Ia. As a simple functional form, we adopt:
\begin{equation}
^{56}\rm{Ni}\left(m\right) = \frac{1}{\exp\left(s\left[m - M_{\rm{Ni}}\right]/M_{\rm{\odot}}\right) + 1 },
\end{equation}
where m is the mass coordinate of the ejecta and M$_{\rm{Ni}}$ is the total $^{56}$Ni mass in M$_{\rm{\odot}}$. The scaling parameter, s, is used to alter the shape of the $^{56}$Ni distribution; smaller values have a more shallow $^{56}$Ni distribution, while larger values produce a distribution that sharply transitions between $^{56}$Ni-rich and $^{56}$Ni-poor regions. We present three values of s (3, 9.7, and 100), representing models with an extended, intermediate, and compact $^{56}$Ni distribution, respectively (see Fig.~\ref{fig:model-profiles}(c)). We have also fixed the total $^{56}$Ni mass to be 0.6 $M_{\rm{\odot}}$ for all models discussed here. Future work will test the effects of varying $^{56}$Ni masses, as well as other forms for the $^{56}$Ni distributions. 

\par

As we wish to test only the effects from different distributions of $^{56}$Ni, we have taken a simplified approach to the ejecta composition. In each zone, $^{56}$Ni constitutes 100\% of the total iron group elements immediately after explosion. This is unlikely to be realised in nature but nevertheless allows us to easily test $^{56}$Ni distributions in our simple parameter study. The outer $\sim$0.1 $M_{\odot}$ of the ejecta is dominated by carbon and oxygen, while the remaining material is intermediate mass elements. Relative abundance ratios are determined by the W7 model \citep{nomoto-w7}. Future work will test other composition arrangements. Figure~\ref{fig:model-profiles} shows the density and $^{56}$Ni profiles discussed and representative composition for one model (v12500\_d0\_n8\_s9.7).

\par

%
%______________________________________________________________

\section{Model light curves}
\label{sec:model_lightcurves}
We perform simulations for the models discussed in \S\ref{sect:construct_models} and presented in Table~\ref{tab:model-params}. We first consider the influence of the $^{56}$Ni distribution for a fixed density profile (\S\ref{sec:56ni-params}) and then the effects of different density profiles (\S\ref{sec:dens-params}).

\subsection{Effects of the $^{56}$Ni distribution for a fixed density profile }
\label{sec:56ni-params}

\begin{figure}
\centering
\includegraphics[width=\columnwidth]{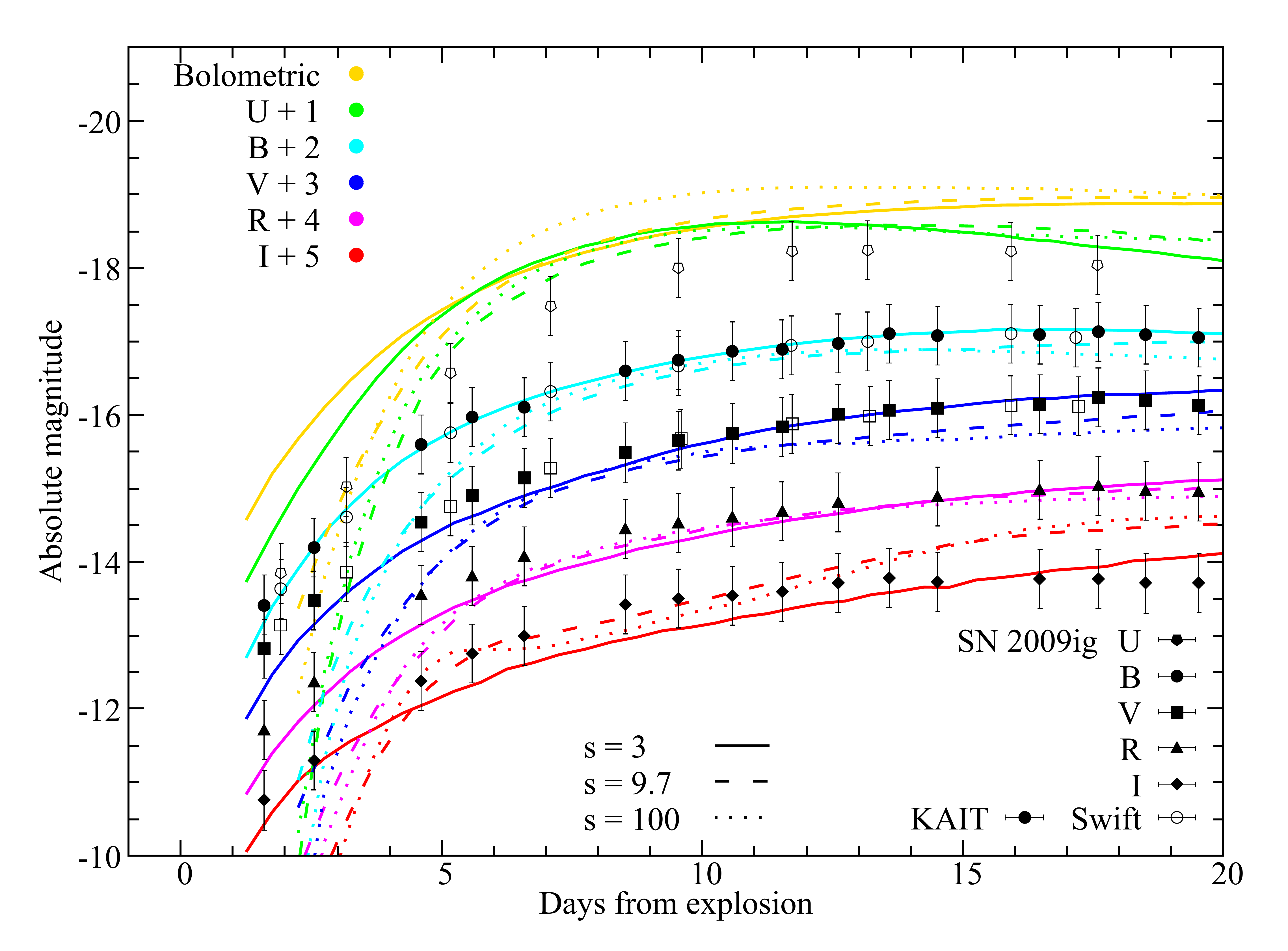}
\caption{Bolometric, Swift $U$, and Johnson $BVRI$ light curves for our v7500\_d0\_n8 density profile, with various scaling parameters (s = 3, 9.7, and 100), and the $UBVRI$ light curves of SN~2009ig. Light curves of SN~2009ig \textbf{are shown} assuming $\mu = 32.6\pm0.4$, negligible host extinction, and an explosion epoch of JD = 2455063.41$\pm$0.08 \citep{foley--2012--09ig}. Filled points represent KAIT light curves, while unfilled points show Swift light curves. We note that for our s = 3 model t$_{\rm{start}}$ = 0.5 days, hence we have omitted the light curve for t \textless 1 days. For s = 9.7 and 100, t$_{\rm{start}}$ = 1.5 days, and we have omitted the light curve for t \textless 2 days in these cases.}
\label{fig:model-lc-09ig}
\centering
\end{figure}

In Fig.~\ref{fig:model-lc-09ig}, we show the light curves for our v7500\_d0\_n8\_s3, 9.7,100 models. Similar to \cite{piro-16}, we show that extended models (e.g. s3) exhibit very different light curves to compact models (e.g. s9.7 and s100). Our s3 model is much brighter than either the s9.7 or s100 models immediately following explosion (by $\gtrsim$2 mag in the $B-$band at three days post-explosion), and shows a shallower rise to maximum. Between three and ten days after explosion, the $B-$band light curve increases by an average rate of 0.30, 0.49, and 0.54 mag~day$^{-1}$ for s3, s9.7, and s100, respectively. In the case of s3, there is a relatively large $^{56}$Ni mass in the outer ejecta - the outer $\sim$50\% of the ejecta mass contains $\sim$20\% of the $^{56}$Ni mass. As this $^{56}$Ni decays to $^{56}$Co, emitted $\gamma$-rays (and subsequent $UVOIR$ photons) experience fewer interactions and escape more easily, hence the light curve is brighter at earlier times. The s9.7 and s100 models have $^{56}$Ni distributions more concentrated towards the ejecta centre - the outer $\sim$50\% of the ejecta contains $\lesssim$5\% of the $^{56}$Ni mass in these cases. Therefore, in these models, emitted light has a higher probability of interaction, due to the larger ejecta mass through which it must travel before escaping, hence the light curves are fainter.

\par

The s3, s9.7, and s100 models also show significant variation in their colour evolution, again for a fixed density profile. This is shown in Fig.~\ref{fig:grey_vs_non-grey} for our v12500\_d0\_n8 models. At very early times, the s3 model shows a bluer $B-V$ colour (by $\lesssim0.3$ mag at three days post-explosion) than s9.7 and s100. During the next few days, the s3 model becomes slightly bluer (by $\lesssim$0.2 mag until approximately one week post-explosion) before gradually becoming redder until the end of our simulation -- at 20 days, the $B-V$ colour of s3 has increased to $\sim$0.75 mag. Overall, the s9.7 and s100 models follow a broadly similar trend, although at different times. At three days, both models show redder colours than s3 ($B-V \sim$0.3 mag) and continually become bluer until approximately two weeks post-explosion -- reaching their most blue colours approximately one week after the s3 model. Following this, both models gradually become redder until the end of our simulations ($B-V\sim$0.3 mag at 20 days).

\par

That s3 is bluer at early times and redder at later times may be understood by considering the effects of $^{56}$Ni heating within the ejecta. The s3 model has a significant amount of $^{56}$Ni present in the outer ejecta that heats its surroundings as it decays to $^{56}$Co. The outer regions are therefore locally heated at all times. The outer regions of the s9.7 and s100 models lack $^{56}$Ni, hence they rely on diffusion of heat from the hotter inner layers and are relatively cool at early times. This produces an initially redder colour than s3, that gradually becomes bluer as light emitted in the inner regions diffuses outwards. As time increases, the outer ejecta become increasingly optically thin, exposing deeper and deeper layers of the ejecta. The more extended $^{56}$Ni distribution in s3 results in less $^{56}$Ni heating of the ejecta interior. Hence the temperature in these regions is lower than in s9.7 and s100, and the colour appears redder.

\subsection{Effects of varying density profiles}
\label{sec:dens-params}

The $^{56}$Ni distribution is affected not only by the scaling parameter s, but also the shape of the density profile (controlled by v$_{\rm{t}}, \delta$, and n). Having shown how the light curves vary for the same density profiles but different scaling parameters, we now discuss the broad features of all of our models, and investigate the effects of each density parameter in turn. 

\par

Our models show a complicated behaviour with varying $^{56}$Ni distributions and density profiles.  
In Table~\ref{tab:model-params} we list the light curve parameters for these models. We fit the light curves from each model with a sixth order polynomial between 10 and 25 days post-explosion. This was chosen simply to produce the best match to all of the model light curves after the initial rising phase but before the end of our simulations. From these model fits, we determine rise times and peak magnitudes in each band. These are given for bolometric, $B-$, and $V-$band light in Table~\ref{tab:model-params}. Table~\ref{tab:model-params} also lists $B-V$ colours at an early time (t = 2.5 days) and at $B-$band maximum. We note that, to determine the colour at 2.5 days, we fit the light curve between two and ten days. 

 \par

The choice of transition velocity significantly affects the shape of the density profile. A higher transition velocity, v$_{\rm{t}}$, results in a larger and less dense inner region, and a smaller and more dense outer region (see Fig.~\ref{fig:model-profiles}). Typically this creates a brighter light curve before maximum light and a shorter rise time. Models with v$_{\rm{t}}$ = 7\,500km~s$^{-1}$ show a median bolometric rise time of 20.1$\pm$1.2 days. For models with v$_{\rm{t}}$ = 12\,500km~s$^{-1}$, the rise time is typically shorter (median $\sim$ 16.7$\pm$0.9 days). Indeed, all v$_{\rm{t}}$ = 7\,500km~s$^{-1}$ models have $V-$band rise times $\geq$23.7 days (most have rise times $\gtrsim$25 days) - higher than the typical rise time for v$_{\rm{t}}$ = 12\,500km~s$^{-1}$ (median $\sim19.7\pm1.2$ days). Higher v$_{\rm{t}}$ will spread out more $^{56}$Ni to higher velocities, therefore light will be allowed to escape more easily in these cases. Similarly, v$_{\rm{t}}$ has a significant effect on the colour evolution. For $\lesssim$3 to 7 days after explosion, higher v$_{\rm{t}}$ models generally appear bluer as they produce more high frequency photons - resulting from the increased density in the outer regions, and hence higher temperature. These regions quickly become optically thin, however, exposing the inner regions. Models with higher v$_{\rm{t}}$ will produce cooler inner regions, appearing redder at later times in both $U-B$ and $B-V$. Despite differences in rise times and colour evolutions, models with different v$_{\rm{t}}$ generally produce similar peak absolute magnitudes (within $\sim$0.2$\pm$0.2~mag).

\par

The effect of a steeper inner density profile (larger $\delta$) is less pronounced at times up to maximum light. A larger $\delta$ shifts mass to lower velocities, causing a decrease in density at higher velocities. This change in density is relatively small everywhere except the very centre of the ejecta (Fig.~\ref{fig:model-profiles}). This generally produces a delay in the rise of the light curve, a slightly longer rise time (by $\lesssim$1.5 days in all but one case), and a small decrease in peak magnitude (within typical photometric uncertainties of $\sim$0.02 for the $B-$ and $V-$band). Higher $\delta$ values may become more significant at later times, as the deepest layers of the ejecta are exposed, but this is beyond the scope of this work. 

\par

For steeper outer density profiles (larger n), again, more of the ejecta mass is shifted to lower velocities. The result is similar to that of a larger $\delta$: the light curve is generally fainter in the optical bands during the pre-maximum phase and shows a slightly longer rise time. The $U-$band, however, is brighter for larger n. This is likely due to the fact that the inner regions are hotter for higher values of n, and the lower density outer region allows photons to escape more easily. Hence, more of the light escapes as higher frequency photons and the other optical band light curves, although fainter, also appear bluer. The effect is less significant in models with higher s values, as most of the $^{56}$Ni is already concentrated in the inner regions for these models. 
\par

We have shown how the model light curves respond to different density and $^{56}$Ni distributions.
Our models broadly reproduce the peak magnitudes and rise times observed in SNe Ia. The exact shape of the light curves and colours, however, may not agree exactly. With an understanding of each parameter, tweaks may be made to produce light curves that better match observations. Model light curves will also be affected by changes in composition, but this will be investigated in future work.

%
%______________________________________________________________

\section{Importance of a non-grey opacity}
\label{sect:grey}

\begin{figure*}
\centering
\includegraphics[scale=0.245]{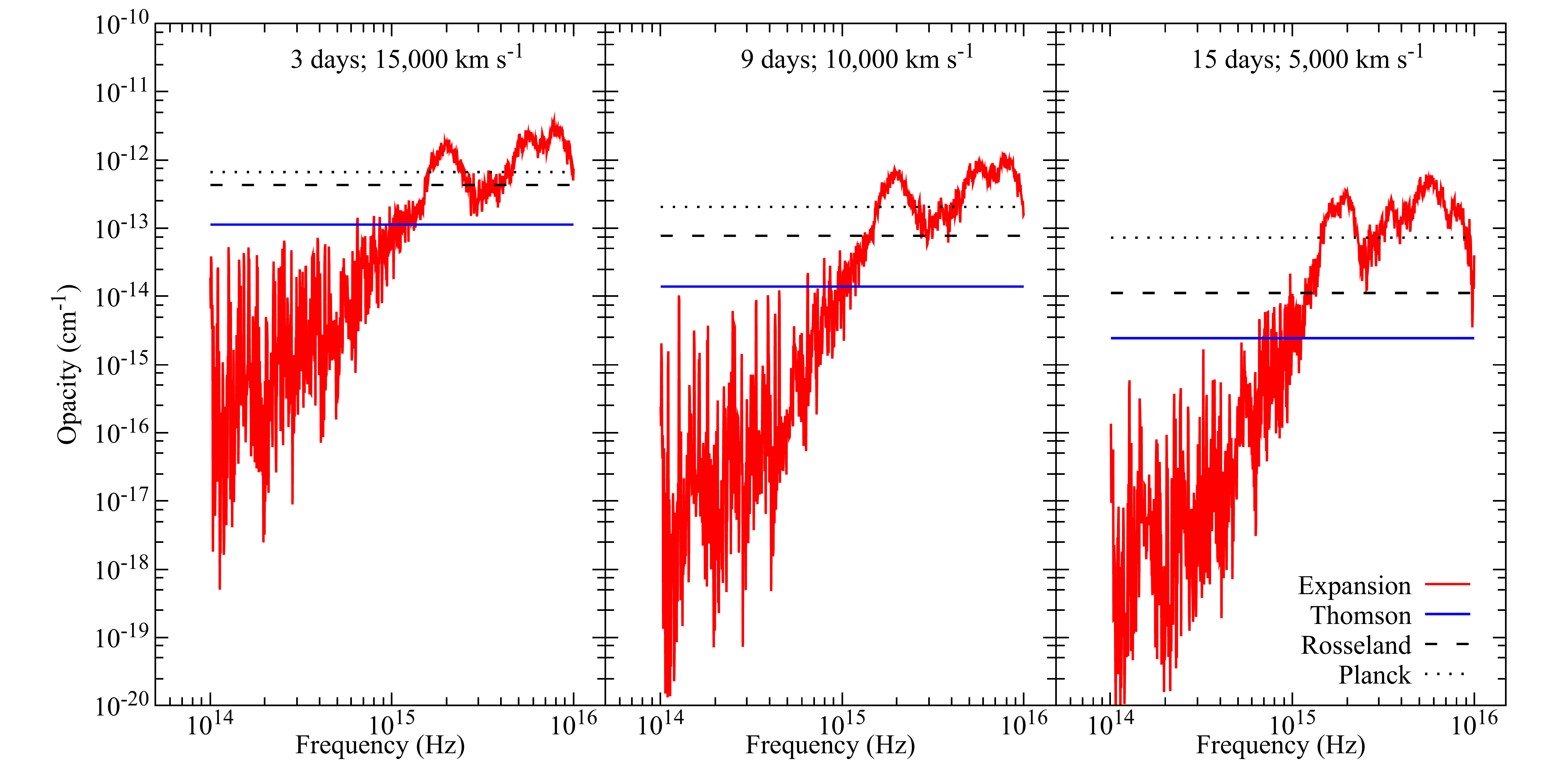}
\caption{Comparison of opacities used in simulations of our v12500\_d0\_n8\_s3 model for three ejecta positions, at three different times. The Rosseland and Planck mean opacities are shown compared to the expansion and Thomson scattering opacities used in the non-grey case.}
\label{fig:grey_vs_non-grey_opacities}
\centering
\end{figure*}

Given the complicated behaviour displayed by our models, here we demonstrate the importance of a frequency-dependent opacity in effectively capturing the evolution of the light curves and colours. We take one density profile (v12500\_d0\_n8) and perform new simulations using a grey, mean opacity for each of the $^{56}$Ni distributions discussed in \S\ref{sect:construct_models} (i.e. extended, intermediate, and compact). 

\par

During each simulation, we calculated expansion and Thomson scattering opacities using TARDIS, as described in \S\ref{subsect:packet-interactions}. We added the additional step of calculating either the Planck or Rosseland mean opacity, which is then used during the next time step. We calculate the Planck mean opacity as:
\begin{equation}
\kappa_P = \frac{\int_0^\infty \kappa_{\rm{Tot}}\left(\nu\right) B_{\nu}\left(T\right) d\nu}{\int_0^\infty B_{\nu}\left(T\right) d\nu} = \frac{\pi}{\sigma T^4} \int_0^\infty \kappa_{\rm{Tot}}\left(\nu\right) B_{\nu}\left(T\right) d\nu,
\end{equation}
while the Rosseland mean opacity is given as:
\begin{equation}
\frac{1}{\kappa_R} = \frac{\int_0^\infty \kappa_{\rm{Tot}}\left(\nu\right)^{-1} \partial B_{\nu}\left(T\right)/\partial T d\nu}{\int_0^\infty \partial B_{\nu}\left(T\right)/ ~\partial T d\nu},
\end{equation}
where B$_{\nu}$(T) is the Planck function. $\kappa_{\rm{Tot}}(\nu)$ is the total opacity, given by $\kappa_{\rm{Tot}}(\nu)$ = $\kappa_{\rm{exp}}(\nu)$ + $\kappa_{\rm{Th}}$, where $\kappa_{\rm{exp}}(\nu)$ is the \cite{eastman-pinto-93} expansion opacity (given in Eqn.~\ref{eqn:exp_opacity}) and $\kappa_{\rm{Th}}$ is the Thomson scattering opacity. The Planck mean opacity is more appropriate for optically thin plasmas (such as the diffuse outer regions of the SN ejecta) and is dominated by frequency intervals with high opacity, while the Rosseland mean opacity is more applicable for optically thick plasmas (such as the inner regions of the ejecta) and is dominated by frequency intervals with low opacity. 

\par

For these simulations, as we use a grey opacity, escaping packets are no longer binned by frequency and only contribute to the observed bolometric luminosity. To convert our bolometric luminosity to observed colours, we find the position of the photosphere ($\tau$ = 2/3), and calculate an effective black-body temperature, following the Stefan-Boltzmann law. We then calculate bolometric corrections for the desired filters using this effective temperature. This is repeated for each point in the light curve.

\par

\begin{figure*}
\centering
\includegraphics[scale=0.295]{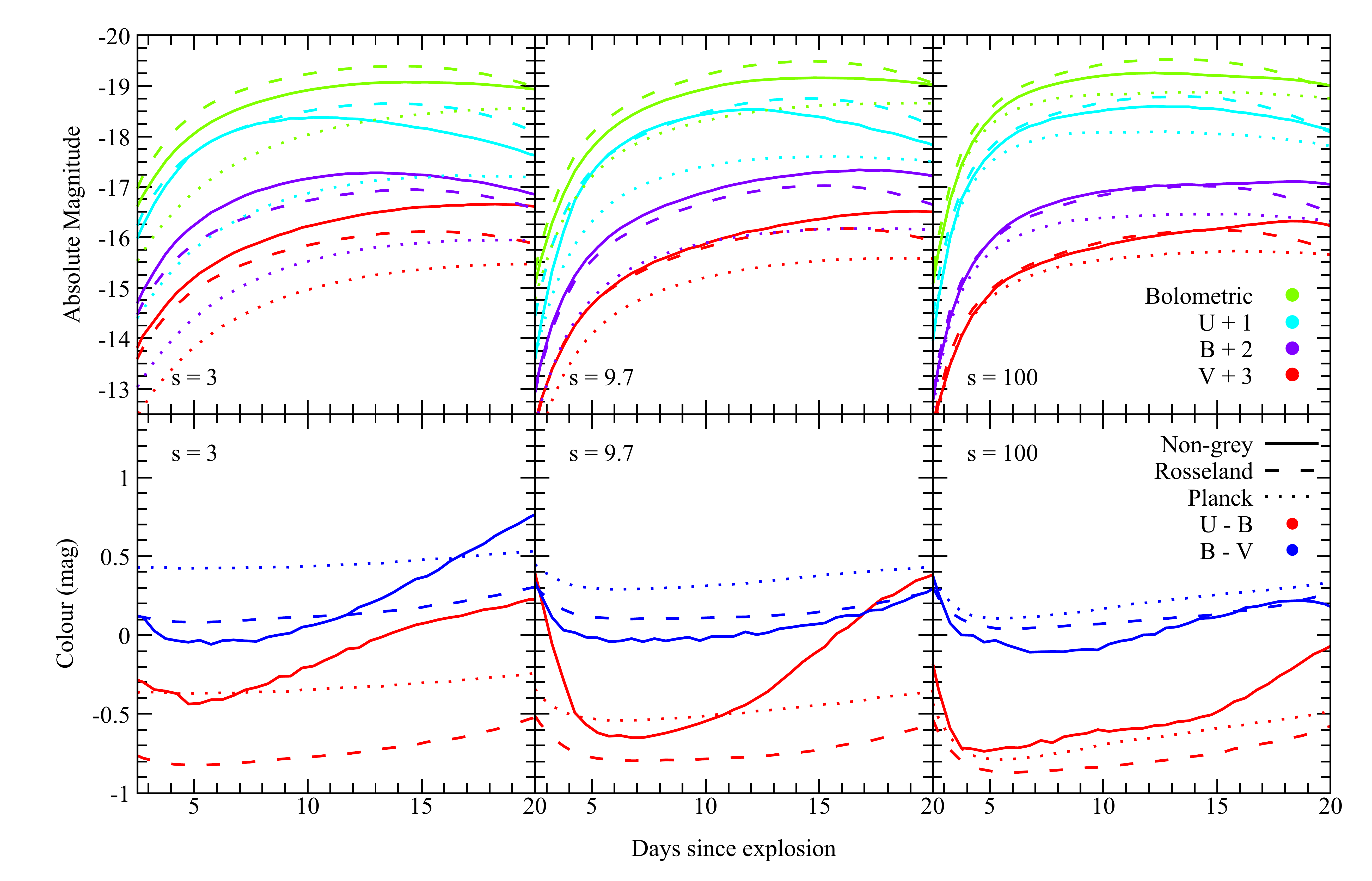}
\caption{Light curves obtained for various opacity treatments. Top: Light curves of v12500\_d0\_n8 models using grey opacities and using our full, non-grey opacity treatment. Light curves have been offset vertically from each other for clarity. Bottom: $B-V$ colours for v12500\_d0\_n8 models using grey opacities and using our full, non-grey opacity treatment.}
\label{fig:grey_vs_non-grey}
\centering
\end{figure*}

Figure~\ref{fig:grey_vs_non-grey_opacities} shows the effect of both approximations on the model opacity for three ejecta velocities at different times. As the Rosseland mean opacity is dominated by low opacity frequency intervals, it typically under-estimates the opacity for UV and near-UV photons ($\nu \gtrsim 10^{15}$ Hz). Conversely, the opacity is over-estimated for optical photons ($\nu \lesssim 10^{15}$ Hz). Figure~\ref{fig:grey_vs_non-grey} demonstrates how the light curves are affected by this approximation, and presents light curves calculated using our full non-grey opacity, and the Planck and Rosseland means. It is clear from Fig.~\ref{fig:grey_vs_non-grey} that the Rosseland mean opacity produces light curves that are overall fainter in optical light, while being brighter in UV and near-UV light. Most photons are injected with UV or near-UV frequencies and therefore experience fewer interactions than in the non-grey opacity case. These photons then escape the ejecta more easily and produce a brighter $U-$band light curve. The optical light curves are directly affected by this also, as fewer photons are reprocessed to lower frequencies. In combination with this is the fact that the optical opacities themselves are also higher, which will further contribute to fainter optical light curves. 

\par

The Planck mean opacity produces an overall similar effect to that of the Rosseland mean opacity. From Fig.~\ref{fig:grey_vs_non-grey_opacities}, the Planck mean opacity is significantly higher than both the Rosseland mean and non-grey opacities for all frequencies (with the exceptions of a narrow frequency window centred around $\nu \sim2\times10^{15}$ Hz and $\nu \textgreater 5\times10^{15}$ Hz). As a consequence, packets experience more interactions, hence fewer photons escape and the light curves are fainter (Fig.~\ref{fig:grey_vs_non-grey}).

\par

From Fig.~\ref{fig:grey_vs_non-grey} it is clear that grey opacities and a colour correction are insufficient to capture the full evolution of the model light curves, and they become increasingly poor at later times and for models with extended $^{56}$Ni distributions. This is unsurprising, given that in models with more extended $^{56}$Ni distributions, there may be a relatively large $^{56}$Ni mass above the photosphere. Assuming these models operate as a black-body with a well defined photosphere is therefore a rather poor approximation. Even in models with $^{56}$Ni concentrated more towards the ejecta centre, the photosphere approximation becomes increasingly poor at later times, as the photosphere recedes into deeper layers of the ejecta where $^{56}$Ni is more prominent.

\par

Having investigated each opacity approximation, we have shown that the Rosseland mean opacity can serve as a moderately good approximation for the early light curves in some regards. However, this grey opacity approximation is not suitable for cases where there is a large source of energy external to the photosphere -- in other words, it has significant limitations when applied to models with extended $^{56}$Ni distributions (at all times) or if applied later than the first few days (all models). Therefore, complete quantification of the effects of the $^{56}$Ni distribution does need a full non-grey radiative transfer treatment.

%
%______________________________________________________________

\section{Relevance of surface $^{56}$Ni to light curve properties}
\label{sec:surface_ni}

\begin{figure*}[!t]
\centering
\includegraphics[scale=0.32]{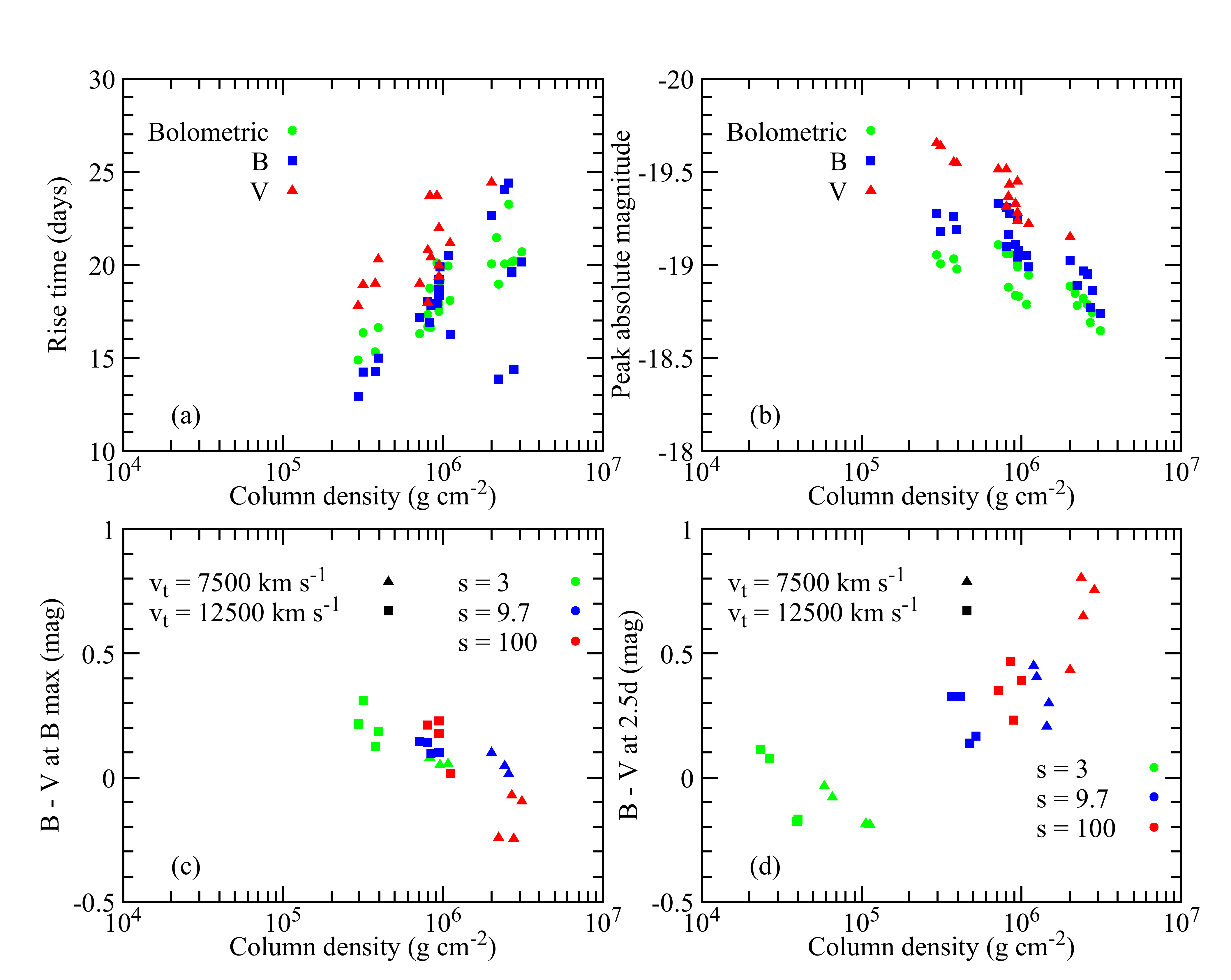}
\caption{Correlations between model column density and light curve parameters. (a) Rise time as a function of column density for bolometric, $B-$, and $V-$band light curves. (b) Peak absolute magnitude as a function of column density for bolometric, $B-$, and $V-$band light curves. (c) $B-V$ colour at time of $B-$band maximum. (d) $B-V$ colour at 2.5\,d after explosion. For (a), (b), and (c) we first find the radius that encloses 90\% of the total $^{56}$Ni mass. We then integrate density over radius (at 10\,000s) outwards from this point to calculate the column density. For (d), we follow the same procedure but instead calculate the column density above the radius at which 99\% of the total $^{56}$Ni mass is enclosed. We find evidence for correlations (of varying significance) between column density, rise time, peak magnitude, and colour. 
}
\label{fig:model-correlations}
\centering
\end{figure*}

Although the total ejecta mass is fixed for each model, the shape of the density profile is controlled by three parameters (v$_{\rm{t}}$, $\delta$, n), with an additional parameter, s, controlling the distribution of $^{56}$Ni within the density profile. We have shown how each parameter affects the light curve, however they may also may be combined into a single parameter, column density. This gives an overall estimation of the amount of $^{56}$Ni present in the outer ejecta, or the proximity of the majority of $^{56}$Ni to the ejecta surface (i.e. higher column densities indicate the majority of the $^{56}$Ni mass is more deeply embedded within the ejecta). Following \cite{noebauer-17}, we investigate correlations between column density (measured at 10$^4$s after explosion) and the light curve properties given in Table~\ref{tab:model-params}. The column density is given by:
\begin{equation}
\eta = \int_{r_{\rm{Ni}}}^{\infty} \rho\left(r\right) dr.
\end{equation}
\cite{noebauer-17} define r$_{\rm{Ni}}$ to be the outermost point at which $^{56}$Ni constitutes 1\% of the ejecta composition. As the mass fraction of $^{56}$Ni does not drop to 1\% in some of our models (see Fig.~\ref{fig:model-profiles}), we instead define r$_{\rm{Ni}}$ as the point at which the majority of $^{56}$Ni is enclosed - in other words, the radius at which 90\% (or 99\%; see Table~\ref{tab:model-params} for further details) of the total $^{56}$Ni mass is contained. Table~\ref{tab:model-rise} gives the column densities measured for our models. In determining the significance of trends, we use the Spearman rank correlation coefficient, $\mathcal{R}_{\rm{S}}$ - a measure of whether any possible monotonic correlation (but not necessarily linear) exists between two variables. We note that $\mathcal{R}_{\rm{S}}$ = $\pm$1 indicates the variables are completely correlated, while zero indicates no monotonic correlation. 

\par

Figure~\ref{fig:model-correlations}(a) shows the rise times for our bolometric, $B-$, and $V-$band light curves as a function of column density. We find evidence for a strong correlation between bolometric rise time and column density, with $\mathcal{R}_{\rm{S}}$ = 0.90 ($p$-value $\sim4\times10^{-9}$). The column density defined here is simply a measure of the amount of surface $^{56}$Ni, therefore it is unsurprising that models with $^{56}$Ni close to the ejecta surface (i.e. the column density is low) have significantly shorter rise times. Correlations in the $B-$ and $V-$bands are less significant, with $\mathcal{R}_{\rm{S}}$ = 0.57 ($p$-value $\sim4\times10^{-3}$) and $\mathcal{R}_{\rm{S}}$ = 0.73 ($p$-value $\sim2\times10^{-3}$), respectively. The strength of the correlation in the $V-$band is likely affected by the fact that a large number of our models peak after or close to the end of the simulations, and hence are not included. While the correlation between column density and rise time is fairly strong in bolometric light, it is less so in the $B-$ and $V-$bands. This would indicate that there are more subtle effects determining the colour light curves than purely the $^{56}$Ni distribution. The v7500\_d0,1\_n8\_s100 models, for example, are noticeable outliers in the $B-$band - their rise times are significantly shorter than models with similar column densities. Hence, we caution against attempts to determine the levels of surface $^{56}$Ni simply by measuring rise times in individual observed bands.

\par

Similarly, we test for correlations between the column density and peak absolute magnitude. For the bolometric, $B-$, and $V-$band light curves we find $\mathcal{R}_{\rm{S}}$ = 0.82 ($p$-value $\sim9\times10^{-7}$), 0.88 ($p$-value $\sim4\times10^{-8}$), and 0.88 ($p$-value $\sim1\times10^{-5}$), respectively. The primary factor in determining the peak brightness of SNe Ia is the total amount of $^{56}$Ni produced during the explosion. Our results indicate that the distribution of $^{56}$Ni itself adds a further complication. Future models with varying $^{56}$Ni masses will allow for an investigation into the degeneracy between the total amount of $^{56}$Ni and its distribution.

\par

Finally, we investigate column density and colour. We find that models with higher levels of surface $^{56}$Ni (low column densities) tend to show redder colours at $B-$band maximum ($\mathcal{R}_{\rm{S}}$ = $-$0.84; $p$-value $\sim3\times10^{-7}$). Conversely, these models produce bluer colours very shortly after explosion ($\mathcal{R}_{\rm{S}}$ = 0.82; $p$-value $\sim9\times10^{-7}$). As in the case for fixed density profiles, these trends demonstrate that models with extended $^{56}$Ni distributions (lower column densities) are typically hotter at early times and cooler at later times. Again, this is a result of the amount of $^{56}$Ni heating being probed at different points within the ejecta.

\begin{table*}
\centering
\caption{Ejecta model rise indices and column densities}\tabularnewline
\label{tab:model-rise}\tabularnewline
\begin{tabular}{llllllll}\hline
\hline
\tabularnewline[-0.25cm]
Model & Bolometric 	& $U$  	& $B$  	& $V$	   	& $R$ 	    & $I$	   	& Column  \tabularnewline
	  &  	  	 	&    	&    	&       	&        	&      		& density \tabularnewline
 	  &   	 	 	&    	&    	&       	&        	&      		& $\eta$ (g~cm$^{-2}$) \tabularnewline
 
\hline
\hline
\tabularnewline[-0.2cm]
v7500\_d0\_n8\_s3		&	1.84	&	2.55	&	1.76	&	1.59	&	\nodata		&	\nodata	 	&	8.32 $\times 10^{5}$	\tabularnewline 
v7500\_d0\_n8\_s9.7		&	2.55	&	3.62	&	2.69	&	2.05	&	2.10	&	1.54 	&	2.01 $\times 10^{6}$	\tabularnewline 
v7500\_d0\_n8\_s100		&	3.02	&	4.03	&	3.34	&	\nodata		&	2.37	&	1.46	&	2.21 $\times 10^{6}$	\tabularnewline 
v7500\_d0\_n12\_s3		&	1.62	&	1.83	&	1.55	&	\nodata		&	\nodata		&	\nodata		&	9.55 $\times 10^{5}$	\tabularnewline 
v7500\_d0\_n12\_s9.7	&	1.98	&	2.39	&	1.89	&	\nodata		&	\nodata		&	\nodata		&	2.43 $\times 10^{6}$	\tabularnewline 
v7500\_d0\_n12\_s100	&	2.63	&	3.16	&	2.68	&	\nodata		&	\nodata		&	\nodata		&	2.67 $\times 10^{6}$	\tabularnewline 
v7500\_d1\_n8\_s3		&	1.72	&	2.39	&	1.67	&	1.60	&	\nodata		&	\nodata	 	&	9.18 $\times 10^{5}$	\tabularnewline 
v7500\_d1\_n8\_s9.7		&	2.48	&	3.39	&	\nodata		&	\nodata		&	2.22	&	1.74	&	2.15 $\times 10^{6}$	\tabularnewline 
v7500\_d1\_n8\_s100		&	2.91	&	3.82	&	3.48	&	\nodata		&	2.57	&	1.42	&	2.77 $\times 10^{6}$	\tabularnewline 
v7500\_d1\_n12\_s3		&	1.55	&	1.79	&	1.50	&	\nodata		&	\nodata		&	\nodata		&	1.08 $\times 10^{6}$	\tabularnewline 
v7500\_d1\_n12\_s9.7	&	2.15	&	2.71	&	2.06	&	\nodata		&	\nodata		&	\nodata		&	2.59 $\times 10^{6}$	\tabularnewline 
v7500\_d1\_n12\_s100	&	2.85	&	3.47	&	2.93	&	\nodata		&	\nodata		&	\nodata		&	3.09 $\times 10^{6}$	\tabularnewline 
v12500\_d0\_n8\_s3		&	1.77	&	2.26	&	1.87	&	1.56	&	1.53	&	1.40 	&	2.95 $\times 10^{5}$	\tabularnewline 
v12500\_d0\_n8\_s9.7	&	2.32	&	3.24	&	2.07	&	1.85	&	2.09	&	2.07 	&	7.15 $\times 10^{5}$	\tabularnewline 
v12500\_d0\_n8\_s100	&	2.55	&	3.17	&	2.77	&	1.99	&	2.27	&	1.67 	&	8.03 $\times 10^{5}$	\tabularnewline 
v12500\_d0\_n12\_s3		&	1.62	&	2.16	&	1.58	&	1.56	&	1.51	&	1.51 	&	3.74 $\times 10^{5}$		\tabularnewline 
v12500\_d0\_n12\_s9.7	&	1.98	&	2.79	&	1.90	&	1.60	&	1.77	&	1.62 	&	8.44 $\times 10^{5}$	\tabularnewline 
v12500\_d0\_n12\_s100	&	2.24	&	3.06	&	2.29	&	1.79	&	1.88	&	1.56 	&	9.48 $\times 10^{5}$	\tabularnewline 
v12500\_d1\_n8\_s3		&	1.65	&	2.06	&	1.81	&	1.52	&	1.50	&	1.42 	&	3.14 $\times 10^{5}$	\tabularnewline 
v12500\_d1\_n8\_s9.7	&	2.25	&	3.08	&	2.09	&	1.86	&	2.06	&	2.16 	&	8.05 $\times 10^{5}$	\tabularnewline 
v12500\_d1\_n8\_s100	&	2.43	&	3.30	&	2.61	&	2.00	&	2.38	&	1.89 	&	9.37 $\times 10^{5}$	\tabularnewline 
v12500\_d1\_n12\_s3		&	1.54	&	1.84	&	1.53	&	1.50	&	1.51	&	1.56 	&	3.95 $\times 10^{5}$		\tabularnewline 
v12500\_d1\_n12\_s9.7	&	2.16	&	2.80	&	2.00	&	1.76	&	1.96	&	1.74 	&	9.49 $\times 10^{5}$	\tabularnewline 
v12500\_d1\_n12\_s100	&	2.42	&	3.43	&	2.52	&	1.89	&	2.14	&	1.60 	&	1.11 $\times 10^{6}$	\tabularnewline 
\hline
\hline
Median	&	2.20(0.44)	&	2.93(0.64)	&	2.06(0.56)	&	1.76(0.18)	&	2.08(0.33)	&	1.58(0.22) 	& \tabularnewline 
\hline
\end{tabular}
\tablefoot{Rise indices as measured from Eqn.~\ref{eqn:rise}. Median values for each filter are also given, along with standard deviations in brackets. The models explored here are not exhaustive and future work will be able to further explore an extended range of parameters. We show the median and standard deviations simply to demonstrate the spread achievable from this specific set of models.
}
\end{table*}

\par

Our analysis shows the complicated sensitivity of the light curves to the parameters describing the ejecta, and general trends among the models. Although we find relatively strong correlations with colour and column density, these do not capture the shape of the light curves. For example, our v7500\_d0\_n12\_s3 and v12500\_d1\_n12\_s3 models show similar colours shortly after explosion, however, their colours and light curve shapes quickly diverge. Therefore, while colours and peak magnitudes, for example, may give a general sense of the level of surface $^{56}$Ni, caution is to be advised if one attempts to quantitatively infer the $^{56}$Ni distribution. Fully characterising the $^{56}$Ni distribution requires comprehensive comparisons with complete model light curves and colours.

%
%______________________________________________________________

\section{Rise index}
\label{sec:rise_index}
Having investigated the light curve parameters, we now quantify the overall shape of the rising phase for our models. We perform fits for the rise index ($z$) in the same manner as \cite{firth--sneia--rise}. For each model, we normalise the light curves using the peak absolute magnitudes from our polynomial fits (Table~\ref{tab:model-params}). We then fit the rising phase using:
\begin{equation}
\label{eqn:rise}
f\left(t\right) = \alpha t^{z},
\end{equation}
where $f$ is flux, $t$ is the time since explosion, $\alpha$ is a normalising factor, and $z$ is the rise index. As in \cite{firth--sneia--rise}, we define the rising phase as times where $f$ \textless ~0.5$f_{\rm{Peak}}$. Table~\ref{tab:model-rise} shows our fitted rise indices in each band, along with the median rise index.

\par

\cite{firth--sneia--rise} show how the rise index of SNe Ia covers a broad range, from $\sim$1.5 to $\sim$3.7 for the PTF $R-$band and the broadband LSQ filter (which covers approximately the range of the SDSS $gr$ filters). Our model rise indices are consistent with those of \cite{firth--sneia--rise}, with fits to bolometric light curves producing rise indices ranging from 1.54 to 3.02. Based on the work of \cite{piro-nakar-2014}, \cite{firth--sneia--rise} argue that the observed distribution of rise indices can be explained by differences in the $^{56}$Ni distribution. Our models support this conclusion and show that various $^{56}$Ni distributions can produce a wide range of rise indices. For a fixed density profile, our models show that an increasingly extended $^{56}$Ni distribution generally produces a more shallow rising light curve (lower rise index). However, the rising behaviour shows additional complexities. For a given value of s, the $^{56}$Ni distribution is fixed in terms of mass coordinate, yet changing the density parameters also affects the rise index. Therefore, although useful, the rise index is not a perfect indicator of the $^{56}$Ni mass depth probed by the light curves: the effects of the density and its overall shape must also be considered.

%
%______________________________________________________________

\section{Comparison with the SN Ia SN~2009ig}
\label{sec:09ig}
In Fig.~\ref{fig:model-lc-09ig}, we show the light curve of a SNe Ia, SN~2009ig \citep{foley--2012--09ig}, and demonstrate how our models may be used to constrain the $^{56}$Ni distribution in SNe Ia. The light curve of SN~2009ig shows remarkably good agreement with our v7500\_d0\_n8\_s3 model light curve, despite the relatively simple, parametrised approach to our model setup (e.g. broken power law density profile, simple composition). It is clear from Fig.~\ref{fig:model-lc-09ig} that the early rise of SN~2009ig is too shallow to result from a compact $^{56}$Ni distribution. 

\par

Despite very good agreement in the $BVR$ light curves, there are notable differences between the model and observed $U-$ and $I-$band light curves. Our model $U-$band light curve shows a shallower rise to maximum than SN~2009ig, and is generally brighter (by between $\sim$0.2 - 0.8 magnitudes). The $U-$band is strongly affected by line blanketing from iron group elements (IGE), and even trace amounts of IGE may have a dramatic effect. We have assumed that $^{56}$Ni is 100\% of the total IGE mass (as a means of testing purely the effects from the $^{56}$Ni distribution), while explosion models typically predict that this is not the case. For example, for the set of deflagration-to-detonation transition models presented by \cite{seitenzahl-2013}, $^{56}$Ni constitutes between $\sim$60 - 90\% of the total IGE mass. Therefore, our models likely underestimate the total IGE mass in SNe Ia, hence the $U-$band light curve in particular appears brighter than what is observed. In addition, trace amounts of IGE may also be present in the progenitor WD, and therefore affect the $U-$band light curves \citep{hoeflich-98,lentz--2000}. Future work will test different compositions for our models. 

\par

SN~2009ig shows a relatively flat $I-$band light curve from approximately seven days after explosion. The model light curve instead shows a smooth rise over the same period. This is likely a consequence of our LTE assumption. The recombination of IGE results in a strong secondary maximum in the NIR light curves of SNe Ia, and a slight `shoulder' in some of the optical bands \cite[e.g. $R$ and, to a much lesser extent, $V$;][]{kasen-06a}. As discussed by \cite{artis}, the assumption of LTE has important implications for this effect. \cite{artis} show that when using a detailed ionisation treatment, as opposed to LTE, the iron group elements in the ejecta remain more highly ionised for longer. This generally results in the recombination of \ion{Fe}{iii} to \ion{Fe}{ii} happening at earlier times in LTE. Hence, the first and secondary maxima blend together, forming a single peak that is broader than is observed in SNe Ia. This could explain why the model $I-$band in particular is most discrepant with SN~2009ig. 

\par

Figure~\ref{fig:model-lc-09ig} shows a comparison of our model light curves to SN~2009ig assuming an explosion epoch of JD = 2455063.41. \cite{foley--2012--09ig} infer this explosion epoch for SN~2009ig following the method of \cite{riess--99b}, where L $\propto$ t$^{2}$. \cite{piro-nakar-2014}, however, infer an explosion epoch 1.6 days earlier (JD = 2455061.8) by fitting the velocity evolution of absorption features (where v $\propto$ t$^{-0.22}$). We compared all of our models to SN~2009ig using both explosion epochs and found best agreement with our v7500\_d0\_n8\_s3 model and the explosion epoch of \cite{foley--2012--09ig}. As an additional test, we fit this model to SN~2009ig by varying the explosion epoch. We find a best match explosion time to be consistent with that of \cite{foley--2012--09ig}, JD = 2455063.34. We stress, however, that the model agreement with SN~2009ig is not perfect, and therefore could affect our explosion time estimate.

%
%______________________________________________________________

\section{Conclusions}
\label{sect:conclusions}

We presented a new Monte Carlo code, purpose built for modelling the early light curves of radioactively driven transients. Light curves computed by our code for the well-studied W7 explosion model \citep{nomoto-w7} are consistent with those from other radiative transfer codes.

\par

We performed an extensive parameter study of the $^{56}$Ni distribution and density profile, and explored their effects on model light curves. Similar to \cite{piro-16} and \cite{noebauer-17}, we find that the light curve is strongly affected by the $^{56}$Ni distribution. For a given density profile, models with $^{56}$Ni extending throughout the ejecta are typically brighter and bluer at earlier times than models in which $^{56}$Ni is embedded deep within the ejecta. The density profile, however, also has significant effects on the model light curves.

\par

We demonstrated the importance of a full radiative transfer treatment through comparisons with models that use grey opacities. We show how a grey (Rosseland) opacity is typically only applicable for times less than approximately one week after explosion and for models that do not have extended $^{56}$Ni distributions, and hence may produce inaccurate estimations of the $^{56}$Ni distribution.

\par

Relations between the amount of surface $^{56}$Ni (or column density) and rise time, peak magnitude, and colour were investigated. We found that while correlations do exist, the scatter is sufficiently large that significant caution must be applied if individual light curve parameters (e.g. rise time to $B-$band maximum) are used to infer $^{56}$Ni distributions. A comprehensive comparison of full colour light curves is necessary to quantify the $^{56}$Ni distribution in individual objects.

\par

Finally, we compared our series of models to observations of the SNe Ia SN~2009ig \citep{foley--2012--09ig}. We find remarkably good agreement with a model that has $^{56}$Ni extending to the outer ejecta, despite the relatively simple approach we have taken with, for example, the composition. It is clear that SN~2009ig is inconsistent with a $^{56}$Ni distribution that is concentrated towards the ejecta centre, and likely had a significant amount of $^{56}$Ni present in the outer ejecta. \cite{piro-nakar-2014} conclude that SN~2009ig must have had a $^{56}$Ni mass fraction, X$_{56}$ of $\sim$0.1 at $\sim0.1 M_{\rm{\odot}}$ from the ejecta surface. All of our models with s = 3 have similar compositions to this (i.e. X$_{56}$ = 0.1 at $\sim0.1 M_{\rm{\odot}}$ from the ejecta surface), however only our v7500\_d0\_n8\_s3 model matches the light curve shape of SN~2009ig. This demonstrates the considerable power of early-phase light curve analysis to constrain a range of ejecta properties (such as the density profile) in addition to the $^{56}$Ni mass depth. Future work will focus on models with varying $^{56}$Ni and ejecta masses, as well as different compositions and more complex $^{56}$Ni distributions.

\par

Our models clearly demonstrate that colour information is necessary to characterise the $^{56}$Ni distribution. In the case of SN~2009ig, our models show a preference for an extended $^{56}$Ni distribution, similar to detonation-to-deflagration and pure deflagration models. Whether other explosion models can induce similar degrees of mixing remains to be seen. Upcoming surveys will discover tens of thousands of SNe Ia. The high cadence and large field of view of the {\it Large Synoptic Survey Telescope} \citep{ivezic--08,lsst--09} for example, means that many of these discoveries will occur shortly after explosion. Comparison to a larger observed sample will demonstrate whether the majority of SNe Ia typically show similar $^{56}$Ni distributions and place greater constraints on the explosion mechanism(s) of this class of supernovae.

%
%______________________________________________________________

\begin{acknowledgements}
We are grateful to U. N\"obauer for providing TARDIS analysis tools and STELLA light curves, L. Shingles for providing W7 model files, and C. Inserra for providing filter sets. 
We thank the anonymous referee for suggestions to improve our manuscript.
This work made use of the Heidelberg Supernova Model Archive (HESMA), https://hesma.h-its.org.
RK and SAS acknowledge support from STFC via grants ST/L000709/1 and ST/P000312/1. WEK was supported by an ESO fellowship.

\end{acknowledgements}

\bibliographystyle{aa}
\bibliography{aanda}

\begin{appendix}

\end{appendix}

\end{document}